\documentclass[aps,amsmath,amssymb,twocolumn,superscriptaddress,citesort,longbibliography, prfluids]{revtex4-2}

\usepackage{graphicx}
\usepackage{dcolumn}
\usepackage{bm}
\usepackage{hyperref}
\hypersetup{
    colorlinks=true,
    linkcolor=blue,        
    filecolor=magenta,     
    urlcolor=cyan,         
    citecolor=purple,         
    pdftitle={SOL for Shell Models Closure - AF},
    pdfpagemode=FullScreen,
}

\newcommand{\thetaColor}{magenta}

\usepackage{xcolor}
\usepackage{float}
\floatstyle{plaintop}
\restylefloat{table}
\makeatletter
\let\newfloat\newfloat@ltx
\makeatother
\usepackage{algorithm}
\usepackage{algpseudocode}
\usepackage{setspace}
\usepackage{lipsum}
\usepackage{subcaption}
\usepackage{booktabs} 

\newcommand{\figsize}{1.0}

\newcommand\void[1]{}

\begin{document}


\title{\newstuff{Solver-in-the-loop approach to closure of shell models of turbulence}}

\author{André Freitas}
\email{andre.freitas@roma2.infn.it}
\affiliation{Dept. Physics and INFN, University of Rome “Tor Vergata”, Italy}
\affiliation{LTCI, Télécom Paris, IP Paris, France}
\author{Kiwon Um}
\affiliation{LTCI, Télécom Paris, IP Paris, France}
\author{Mathieu Desbrun}
\affiliation{Inria and École Polytechnique, IP Paris, France}
\author{Michele Buzzicotti}
\affiliation{Dept. Physics and INFN, University of Rome “Tor Vergata”, Italy}
\author{Luca Biferale}
\affiliation{Dept. Physics and INFN, University of Rome “Tor Vergata”, Italy}

\def\LB#1{{\textcolor{black}{#1}}}
\def\BL#1{{\textcolor{blue}{#1}}}
\def\m@#1{{\textcolor{black}{#1}}}
\def\ku#1{{\textcolor{black}{#1}}}
\def\newstuff#1{{\textcolor{black}{#1}}}

\date{\today}

\begin{abstract}
This work studies an {\it a posteriori} data-driven approach (known as {\it solver-in-the-loop}) for sub-grid modeling of \newstuff{ a shell model for turbulence. This approach takes advantage of the} {\it differentiable physics} \newstuff{paradigm of deep learning, allowing a neural network} model to interact with the differential equation solver over time during the training process. The closure model is, then, naturally exposed to {\it equations-informed} input distributions by accounting for prior corrections over the temporal evolution in training. Such a characteristic makes this approach depart from the conventional {\it a priori} instantaneous training paradigm and often leads to a more accurate and stable closure model. Our study demonstrates that the closure learned via this \emph{a posteriori} approach is able to reproduce high-order statistical moments of interest \newstuff{ also } in closures of high Reynolds number turbulence. 
Moreover, we investigate the performance of the learned model by experimenting with the effect of unrolling in time, which has remained for the most part unexplored in the literature. Finally, we discuss potential extensions of this approach to Navier-Stokes equations.

\end{abstract}

\maketitle


\section{Introduction}
\label{sec:intro}

Three-dimensional turbulence is a complex, multiscale phenomenon that arises when the nonlinear transport terms in the Navier-Stokes (NS) equations dominate over viscous damping. The behavior of turbulent flows is governed by the Reynolds number, \(\text{Re} = u_0 l_0/\nu\), where \(u_0\) represents the characteristic velocity, \(l_0\) the typical length scale, and \(\nu\) the kinematic viscosity. At high \(\text{Re}\), turbulence exhibits a range of non-trivial behaviours, including non-Gaussian statistics and intermittent dynamics \cite{frisch}. In 3D turbulence, there is a nonlinear energy cascade, from large to small scales, where it is eventually dissipated through viscous friction \cite{cascades}.

Accurately resolving 3D turbulence is extremely computationally expensive. The degrees of freedom (DOF) scale as a power law of the Reynolds number, $\#_{DOF} \propto Re^{9/4}$, so studying extremely high Reynolds numbers numerically through \newstuff{primitive} Navier-Stokes (NS) equations is often not possible. \newstuff{Modeling is needed.}
A central challenge in turbulence modeling, which has attracted much interest from both theoretical and applied researchers, is Large Eddy Simulation (LES) subgrid scale (SGS) modeling \cite{les_review, lesieur2005, pope2001, sagaut2006}. LES reduces the degrees of freedom encountered in a fully resolved simulation by placing a filter at a certain wavenumber, \newstuff{ $k_c$, and only resolving for $k <k_c$}. This means that the so-called subgrid scales, $k>k_c$, need to be modeled. 
\newstuff{In contrast to what happens in other PDEs set-ups for fluids, such as, e.g., 1D Kuramoto-Sivashinsky equations, 1D Burgers equations, and 2D Navier-Stokes equations, modeling turbulence in 3D is theoretically more challenging because of the strong chaotic, out-of-equilibrium and non-Gaussian nature of high wavenumbers, sub-grid statistics, resulting in multifractal energy dissipation, and extreme subgrid energy transfer fluctuations \cite{frisch}. Furthermore, from a more theoretical and fundamental point of view, the presence of anomalous scaling laws implies a breaking of self-similarity and the existence of a nontrivial dependency of the sub-grid model from $k_c$ \cite{AAM21, Mailybaev2022}. In many applied cases, the cutoff wavenumber cannot be fixed and must be varied ({\it increased}) to improve fidelity of the resolved scale behavior. As a result, a comprehensive theoretical framework defining the statistical properties of the subgrid scale model in 3D turbulence is still missing.}

\newstuff{In this paper, we focus on one specific theoretical aspect of the LES approach, connected to the sub-grid scale anomalous statistical behavior. In order to do that, we need to study the effects of modeling when the $k_c$ falls well inside the inertial range, and the nonlinear energy transfer is strongly non-Gaussian. In contrast to the more established phenomenology-based models \cite{les_review}, we will use a Machine Learning closure, inspired by the complexity of the modeling task and by recent promising results \cite{rm18, rm19, beck1, frezat, marl, mkurz}. The main goal is to attack with high accuracy questions connected to the fidelity of the model to reproduce extreme SGS energy transfer events. No model is perfect, and one expects that extreme rare events are more sensitive to biases. The need to have high $k_c$ (to observe non-Gaussian fluctuations) and very large statistics (for the data-driven approach) makes this study impossible in 3D turbulence, where most of the Machine Learning LES are limited to very small resolution (up to $128^3$ or $256^3$) and, consequently, by a very small departure from quasi Gaussian statistics.}  The only alternative framework where to study these questions is using shell models of turbulence, where only a few degrees of freedom are preserved for a set of logarithmically equispaced wavenumbers, $k_n = k_0 \lambda^n$, where $\lambda=2$ ususally \cite{biferale2003shell}.  Models such as the Sabra model \cite{LvovUnknownTitle1998} have successfully replicated key statistical properties of turbulence, including intermittency, strongly non-Gaussian fluctuations, and anomalous scaling exponents. 
\newstuff{Shell models have been successfully used to study statistical properties of many turbulent fluid configurations, including rotating turbulence \cite{ROT_SM},  thermal convection \cite{THERMAL_SM}, superfluids \cite{SUPERFLUID_SM}, MHD turbulence \cite{MHD_SM}, helical turbulence \cite{HELIC_SM}, and passive scalars \cite{PASSIVE_SC_SM}, to cite just a few. Shell models have also been used to study fundamental properties of NS equations, connected to spontaneous stochasticity \cite{Mailybaev_2016}, effects of thermal noise \cite{bandak24}, existence of solutions \cite{reg_2007}, instantons \cite{instanton}, and many more.}


\newstuff{In this paper, we develop a Deep Learning based SGS closure for shell models of turbulence. Our approach employs an a posteriori training technique known as \emph{solver-in-the-loop}.}
This method incorporates a differentiable solver for the governing equations of a physical system directly into the learning process of a deep neural network tasked with learning the closure. We demonstrate that this approach yields closures that are more stable and perform better than those trained using the traditional static \emph{a priori and instantaneous} paradigm. Additionally, we investigate the concept of the ideal \emph{time in the loop}, a critical aspect that is often overlooked in the literature employing this methodology, and attempt to relate it to a relevant physical quantity.

In \autoref{sec:related_work}, we review prior research on subgrid-scale modeling in LES, emphasizing machine learning closures. We pay particular attention to studies involving shell models and those utilizing differentiable solvers or unrolled training. In \autoref{sec:closure}, we discuss the closure of turbulence shell models within the LES framework and describe our solver-in-the-loop approach to closure in detail. In \autoref{sec:results}, we present and discuss the outputs of our trained models. Finally, \autoref{sec:conclusion} summarizes our findings and outlines potential future research directions.

\section{Related Work}
\label{sec:related_work}
Machine learning, and deep learning in particular, has seen wide adoption in fluid dynamics, as highlighted in several comprehensive reviews \cite{vinuesa2022enhancing}. Generally, machine learning is applied in fluid dynamics either to fully replace a complex system with a surrogate model or to augment existing models by addressing unresolved scales or processes. LES closure falls into the latter category and has drawn significant interest from researchers. Recent studies have explored various approaches, including deep learning \cite{rm18, rm19, beck1, frezat, Cho_Park_Choi_2024} as well as multi-agent and deep reinforcement learning \cite{marl, mkurz}. \newstuff{For a detailed perspective on data-driven turbulence closure, readers are referred to the review by Duraisamy \cite{prf_perspective}.}
\newstuff{State-of-the-art ML tools are not yet able to tackle LES models for highly turbulent flows in the regime where the cutoff wavenumber is high enough to see the strong departure from Gaussianity. This is because of a combination of lack of computational power and/or accuracy, and lack of training data. These questions can be addressed in a quantitative way only in shell models, as of now. }
One of the first contributions of LES closure in the context of shell models of turbulence comes from Biferale et al. \cite{BiferaleUnknownTitle2017}, who developed a theoretical framework to define an optimal subgrid closure. This phenomological based closure stands as a good comparison basis for new approaches. More recently, there has been a noticeable shift toward data-driven techniques. Ortali et al. \cite{OrtaliUnknownTitle2022} made important progress by using a deep recurrent neural network integrated within the time integrator scheme to close the system. Their approach yielded excellent results, especially in capturing both Eulerian and Lagrangian statistics. Another interesting approach is by Domingues Lemos et al. \cite{DominguesLemosUnknownTitle2024}, who used a probabilistic method, specifically a mixture of Gaussians, to close the system. This added a new layer of complexity to LES closure strategies by taking into account the inherent probabilistic nature of the closure.

Among these approaches, Ortali et al.'s method is particularly interesting for us because it achieved the best results and it is the only one based on deep learning. Since they used an architecture with a memory component, they were able to effectively capture the time history effects in the closure. However, they used an \emph{a priori} training approach and, as such, they did not fully account for the compounding effects of model errors over time. Addressing this issue would require unrolling the training process over time. 

The concept of unrolling training in time with differentiable solvers was introduced by Um et al. in 2020 \cite{um2020sol}, under the term \emph{solver-in-the-loop}, \ku{particularly for correcting errors of numerical solvers}. This innovative approach allows for a NN to interact with a differential equation solver for many time steps before performing backpropagation, exposing the NN to (more) correct input distributions, therefore improving the performance of the model when faced with the common distribution/data shift seen in the deployment of these kind of autoregressive models. A key advantage of this method is its reliance on automatic differentiation (AD) frameworks when developing the solver, which allow the gradients to also flow through the solver during backpropagation, leading to more precise unrolled gradients. Writing physical solvers using the AD framework is what is now commonly referred to as \emph{differentiable physics}. More recently, List et al. \cite{list2024temporal} studied extensively the benefits of unrolling in time during training compared to a static instantaneous approach (\emph{a priori} training), as well as the benefits of differentiability in the solver.

Another way to look at the benefits of different training schemes as well as different architectures is through the lens of inductive biases. In machine learning applied to science, models span a spectrum from those that rely almost entirely on data to those heavily informed by physical principles. At one end, fully-connected networks with a non-physics based loss function exemplify purely data-driven approaches, learning patterns directly from data without any built-in assumptions about the underlying system. Moving along the spectrum, convolutional networks \cite{Fukushima_1980} add some inductive biases, such as the assumption of locality and translation invariance, which are particularly effective in image processing. Further along, equivariant networks incorporate symmetries specific to the problem, like rotational symmetry, making them more specialized and efficient. Neural ordinary differential equations push this further by integrating differential equations into the model, embedding a continuous-time understanding of dynamics. Finally, at the most inductive end, models based on \emph{solver-in-the-loop} approach or physics informed NNs, are tightly constrained by well-established physical laws. These models not only learn from data but also ensure that their predictions adhere to known physical principles, making them particularly valuable for complex scientific problems where adherence to physical laws is important. Moreover, they are able to learn with less data than purely data-driven models and tend to generalize better. 

\newstuff{Other researchers have explored the use of differentiable solvers in combination with DL for LES closure. Notably, Sirignano et al. \cite{SirignanoUnknownTitle2020} applied this approach to 3D Homogeneous Isotropic Turbulence (HIT) (at resolution $64^3$), while Shankar et al. \cite{shankar_burgers, shankar_hit} utilized it for the Burgers equation (at resolutions $64-512$)  and 2D HIT (at resolution $64^2$). These efforts do not, alas, extend to very high Reynolds numbers nor address the intense and multi-fractal non-Gaussian statistics typical of real turbulence (2D NSE in the forward enstrophy regime are even globally smooth).  High Reynolds number turbulence presents unique challenges, and it is in this context that shell models become particularly valuable, offering a more tractable framework to study this phenomenon. This is where we believe a research gap exists, and our work aims to address this gap.}

\begin{figure}[htb]
    \centering
    \includegraphics[width=.95\linewidth]{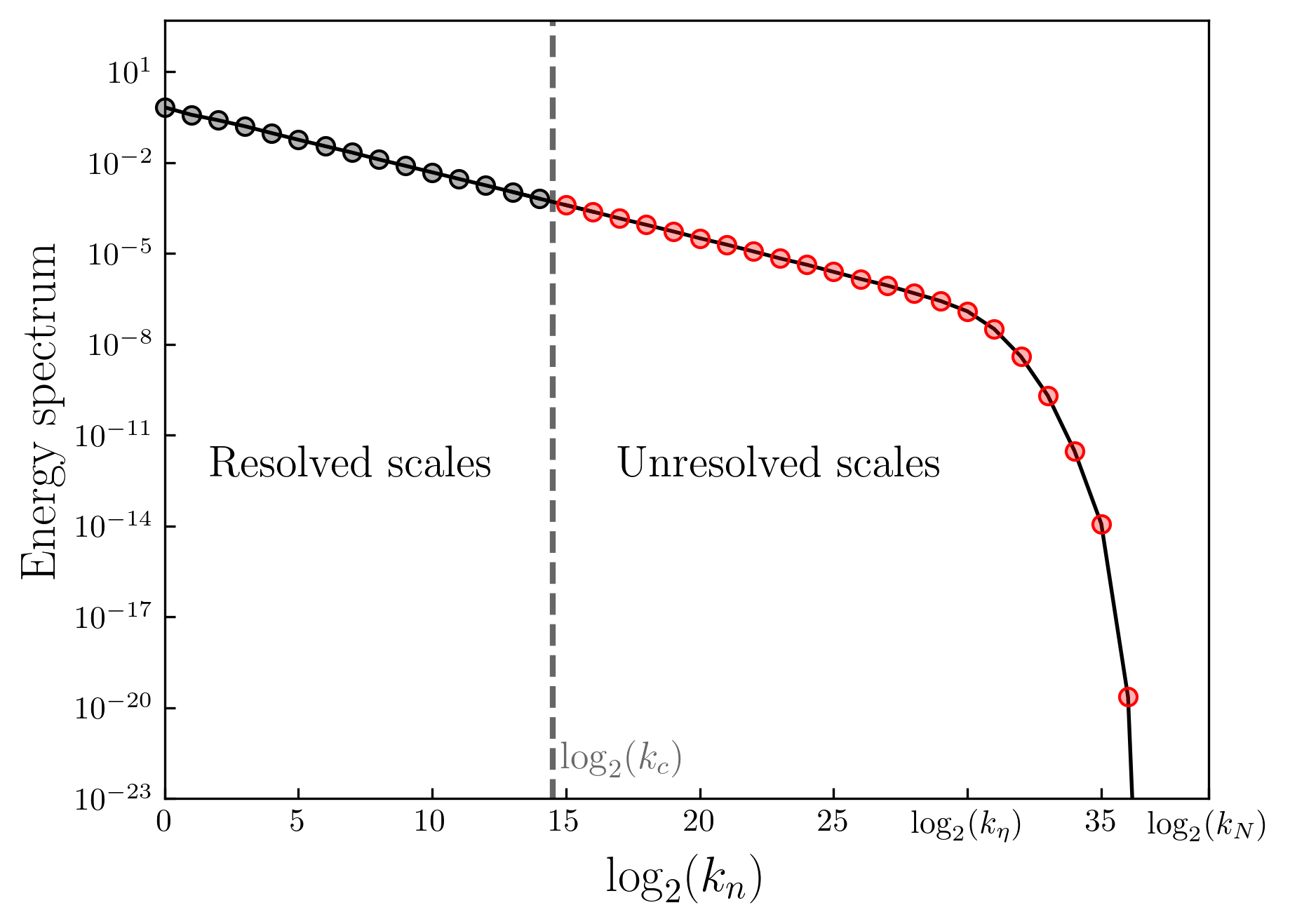}
    \vspace*{-2mm}
    \caption{Energy spectrum showing the large eddy simulation modeling problem. A cutoff is placed in the inertial range, in our case $N_c = 14$. The scales prior to the cutoff are resolved, whereas the one after are unresolved. The influence of the unresolved scales on the resolved ones needs to be modeled. In the case of the Sabra shell model of turbulence, which only has nearest and second nearest neighbor interactions, this means the shells $N_c + 1$ and $N_c + 2$ require modeling.\vspace*{-4mm}}
    \label{fig:ek2}
\end{figure}

\section{Shell Models closure}\label{sec:closure}

Shell models mimic the dynamics of the energy cascade in three dimensional homogeneous isotropic turbulence via a system of coupled non-linear complex-valued  ODEs describing the evolution of the velocity field on a set of wavenuber $k_n$ logarithmically equispaced. In this work, we consider the Sabra model \cite{LvovUnknownTitle1998}, for which the governing equations are:
\begin{align}
    \frac{d u_n}{d t} = i \Big(&a k_{n+1} u_{n+2} u_{n+1}^* + b k_n u_{n+1} u_{n-1}^* \notag \\
    &\quad - c k_{n-1} u_{n-1} u_{n-2}\Big) - \nu k_n^2 u_n + f_n\,,
\end{align}
where $n = 0, \ldots, N$, $k_n = 2^n$,  and $u_n \in \mathbb{C}$. Looking at the right-hand side we can see that, similarly to the NSE in Fourier space, we have a non-linear convective term (which similarly to NSE defines the coupling among wavenumbers; in the shell models only two-away neighbouring interactions are considered) which is the trigger of the energy cascade mechanism, a quadratic dissipative term that dissipates energy at small scales and a forcing term which injects energy at the larger scales.

In a fully resolved system, the number of shells $N$ is determined by the physics of the system. For a higher Reynolds number, the dissipative Kolmogorov length scale, $k_\eta$,  will be at a large wavenumber and as such we have to consider enough shell to resolve it, $k_N > k_\eta$, see \autoref{fig:ek2}. The LES formulation in shells models is similar to a Galerkin Fourier truncation where we consider shells only up until  the cutoff wavenumber $k_{N_c}$ defined by the cutoff shell $N_{c}$ where $N_{c}\!\ll\! N$ and it is usually somewhere in the inertial range. In order to close this reduced model, we need to provide a model for the  two shells right after the cutoff $u_{N_{c}+1}$ and $u_{N_{c}+2}$.
This is depicted below and can be visualized in \autoref{fig:ek2}. We denote the fully resolved model as $u$, while the LES model is represented by $\tilde{u}$.
\[
\begin{array}{c}
\hspace{-2.8cm}
\text{\begin{tabular}{c}
    \textcolor{red}{Fully} \\ \textcolor{red}{Resolved} \\ \textcolor{red}{Model}
\end{tabular}}
\hspace{0.4cm} 
\begin{cases}
    u_{-1} = u_{-2} = 0 \\
    u_{N + 1} = u_{N + 2} = 0
\end{cases} \\[25pt] 
\hspace{0cm}
\text{\begin{tabular}{c}
    \textcolor[HTML]{228B22}{Large} \\ 
    \textcolor[HTML]{228B22}{Eddy} \\ 
    \textcolor[HTML]{228B22}{Simulation}
\end{tabular}}
\hspace{0.2cm}
\begin{cases}
    \tilde{u}_{-1} = \tilde{u}_{-2} = 0 \\
    \tilde{u}_{N_c + 1} = \text{unknown, requires modeling} \\
    \tilde{u}_{N_c + 2} = \text{unknown, requires modeling}
\end{cases}
\end{array}
\]

Now, we will introduce our LES-NN model as well as the basis of comparison, the Ground Truth (GT). The GT is simply the integration of the fully resolved system to generate training and testing data. This system is integrated over a long period to ensure that sufficient data is available to accurately compute the high-order moments of interest. Both the GT and LES-NN are integrated in time using a fourth-order Runge-Kutta (RK4) scheme with the viscous term integrated explicitly. However, different time steps are of course used: the GT is integrated with one much smaller than the LES-NN model to ensure the Kolmogorov scale ($N_{\eta}$) is resolved.

\begin{figure*}[t]
    \centering
    \includegraphics[width=1.\linewidth]{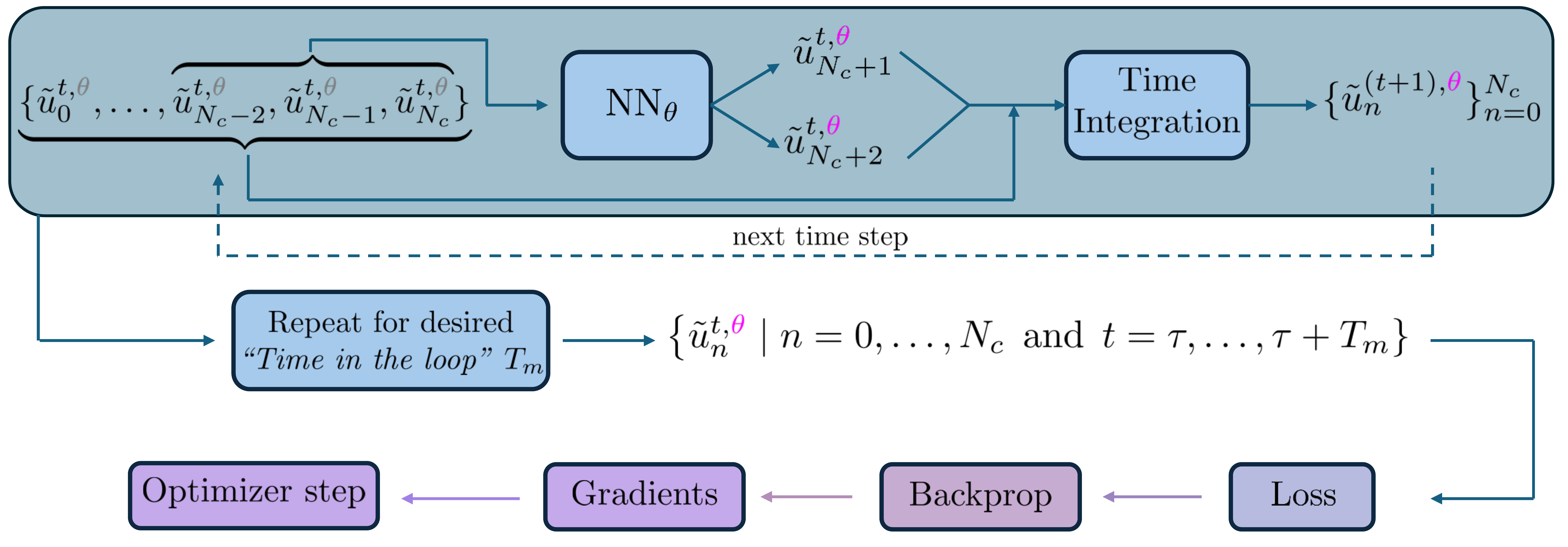}
    \caption{Schematic representation of the LES-NN closure, illustrating how the neural network provides the necessary shells to close the system. Starting at time $t$, the NN takes as input the last three shells before the cutoff (this locally fixes the flux) $\{\tilde{u}_{N_c - 2}^{t, \textcolor{gray}{\theta}}, \tilde{u}_{N_c - 1}^{t, \textcolor{gray}{\theta}}, \tilde{u}_{N_c}^{t, \textcolor{gray}{\theta}}\}$ and outputs the two shells after $\{\tilde{u}_{N_c + 1}^{t, \textcolor{magenta}{\theta}}, \tilde{u}_{N_c + 2}^{t, \textcolor{magenta}{\theta}}\}$ ($\textcolor{gray}{\theta}$ denotes an implicit relation with the NN , whereas $\textcolor{magenta}{\theta}$ denotes an explicit one) . This is enough to close the governing equations and evolve them in time to obtain the new state space at time. This process is repeated for a desired \emph{time in the loop}. The resulting velocity field will be used to computed the loss (mean squared error between the prediction and the ground truth).  Backpropagation is applied to compute gradients, followed by an optimization step to update the NN.}
    \label{fig:sketch}
\end{figure*}

While both systems have the same time integration method, the shell models are different. The GT resolves the $\{u_0, \ldots, u_N\}$ using the governing equations. The LES-NN uses the reduced solver that resolved $\{\tilde{u}_0, \ldots, \tilde{u}_{N_{c}}\}$ using the governing equations and then a neural network at each time step estimates $\tilde{u}_{N_{c+1}}$ and $\tilde{u}_{N_{c+2}}$, therefore closing the system. This is shown in \autoref{fig:sketch}. As input to the neural network, we provide the three shells preceding the cutoff, which is sufficient to close the flux locally. Using fewer shells results in significantly poorer performance, while including more shells offers no noticeable improvement.

One implementation is purposely agnostic to the time integrator used: we integrate the missing terms from the governing equations for $\tilde{u}_{N_{c}-1}$ and $\tilde{u}_{N_{c}-2}$ explicitly as shown in \autoref{eq:ncut_1} and \autoref{eq:ncut}. Terms with superscript $^{\textcolor{\thetaColor}{\theta}}$ are the outputs of the NN, while grey text represents an implicit relation with the NN.

\begin{widetext}
\begin{equation}
    \frac{d \tilde{u}_{N_{c}-1}^{\textcolor{gray}{\theta}}}{d t} = i \left(\underbrace{a k_{N_{c}}  \tilde{u}_{N_{c}+1}^{\textcolor{\thetaColor}{\theta}} \tilde{u}_{N_{c}}^*}_{\text{Integrated explicitly}} + \underbrace{b k_{N_{c}-1} \tilde{u}_{N_{c}} \tilde{u}_{N_{c}-2}^* - c k_{N_{c}-2} \tilde{u}_{N_{c}-2} \tilde{u}_{N_{c}-3}}_{\text{Integrated with RK4}}\right) - \nu k_{N_{c}-1}^2 \tilde{u}_{N_{c}-1} 
    \label{eq:ncut_1}
\end{equation}
\begin{equation}
    \frac{d \tilde{u}_{N_{c}}^{\textcolor{gray}{\theta}}}{d t} = i \left(\underbrace{a k_{N_{c}+1} \tilde{u}_{N_{c}+2}^{\textcolor{\thetaColor}{\theta}} \tilde{u}_{N_{c}+1}^{*, {\textcolor{\thetaColor}{\theta}}} + b k_{N_{c}} \tilde{u}_{N_{c}+1}^{\textcolor{\thetaColor}{\theta}} \tilde{u}_{N_{c}-1}^*}_{\text{Integrated explicitly}} - \underbrace{c k_{N_{c}-1} \tilde{u}_{N_{c}-1} \tilde{u}_{N_{c}-2}}_{\text{Integrated with RK4}}\right) - \nu k_{N_{c}}^2 \tilde{u}_{N_{c}} 
    \label{eq:ncut}
\end{equation}
\end{widetext}

\autoref{alg:training} shows the training loop function. For ease of understanding, the shells between the cutoff are shown as $\tilde{u}^{<}$ and the ones after as $\tilde{u}^{>}$. As it can be seen, the gradients are also being propagated through the solver operations (\texttt{RK4}, which calls the rest of the solver functions). It is also possible to perform unrolled training in the case where the solver is not differentiable, but then one needs to either stop the gradient flow during backpropagation whenever the solver is called (which in the end will lead to worse quality gradients) or to provide by hand the AD primitives. By having a differentiable solver, we are able to bypass these two disadvantages and leave all of the hard work to the AD framework --- the obvious downside of this approach is having to (re)write the solver in an AD framework, which also comes with a few caveats compared to regular non-AD framework programming. In one training iteration, we evolve the system for \texttt{msteps}, which is a hyperparameter. This represents the time that we evolve the system before backpropagating the gradients, i.e., before updating the NN weights.

\begin{algorithm}
\caption{Training Loop Algorithm (a single training iteration)}
\begin{algorithmic}[1]
  \State Initialize Gradient Tape
    \State $\tilde{u} \gets \tilde{u}_0  $\Comment{batch of ICs selected randomly from dataset}
    \For{$t = 0$ to $msteps-1$}
        \State $\tilde{u}^{>, \textcolor{\thetaColor}{\theta}}_t \gets \text{NN}_{\theta}(\tilde{u}^{<}_t)$ 
        \State $\tilde{u}^{<}_{t+1} \gets \text{RK4}(\tilde{u}^{<}_t)$ 
        \State $\mathcal{C}_{N_{c}-1} \gets \Delta \tilde t i(a k_{N_{c}} \tilde{u}_{N_{c}+1}^{\textcolor{\thetaColor}{\theta}} \tilde{u}_{N_{c}}^*)$ 
        \State $\mathcal{C}_{N_{c}} \gets \Delta \tilde t i(a k_{N_{c}+1} \tilde{u}_{N_{c}+2}^{\textcolor{\thetaColor}{\theta}} \tilde{u}_{N_{c}+1}^{*, {\textcolor{\thetaColor}{\theta}}} + b k_{N_{c}} \tilde{u}_{N_{c}+1}^{\textcolor{\thetaColor}{\theta}} \tilde{u}_{N_{c}-1}^*)$
        \State $\mathcal{C} \gets \text{concatenate}(\mathcal{C}_{N_{c}+1}, \mathcal{C}_{N_{c}+2})$
        \State $\tilde{u}^< _{t+1} \gets \tilde{u}^<_{t+1} + \mathcal{C}$ 
    \EndFor
    \State Compute Loss
    \State Compute Gradients
    \State Apply Gradients
\end{algorithmic}
\label{alg:training}
\end{algorithm}

The architecture used for our neural network is the Multi-Layer Perceptron (MLP) \cite{nns} with REctified Linear Unit (ReLU) as the activation function. The number of trainable parameters used in the MLP varied during our studies between $1\cdot10^5$ and $4\cdot10^5$, with the latter used for the results presented here. The loss used is the Mean Square Error (MSE) between the prediction of the reduced system LES-NN, $\tilde{u}$, and the ground truth, $u$, \emph{i.e.}, 
\begin{equation}
     \mathcal{L} = \frac{1}{N_{\text{Loss}}} \sum_{n=1}^{N_{\text{Loss}}}  \frac{\|u - \tilde{u}\|_{bs,T_m}^2} {\sqrt{\|u\|_{bs,T_m}^2} \, \sqrt{\|\tilde{u}\|_{bs,T_m}^2} } \,,
     \label{eq:loss}
\end{equation}
where $\|u\|_{bs,T_m}^2 = \sum_{b=1}^{bs} \sum_{t=\tau_b}^{\tau_b + T_m} |u_{n,t,b}|^2$,
$T_m$ denotes the time in the loop, $bs$ the batch size and $N_{loss}$ is the number of shells considered in the loss function, which in our case is equal to six and these are the shells before the cutoff.

\autoref{tab:params} shows the parameters used in the numerical experiments shown in the following section. Regarding the forcing, the first two shells are forced constantly in time with the magnitudes $f_0 = \epsilon$ and $f_1 = 0.7\epsilon$. This forcing ensures zero helicity flux \cite{LvovUnknownTitle1998}.

\begin{table}[b]
    \centering
    \begin{tabular}{llp{4cm}}
        \toprule
        \textbf{Parameter} & \textbf{Value} & \textbf{Description} \\
        \midrule
        $\nu$ & $1 \times 10^{-12}$ & viscosity \\
        $\text{Re}$ & $\approx 10^{12}$ & Reynolds number \\
        $\epsilon$ & 0.5 & forcing \\
        $N$ & $40$ & number of shells \\
        $N_\eta$ & $30$ & Kolmogorov scale \\
        $N_\text{c}$ & $14$ & subgrid cutoff scale \\
        $\tau_0$ & $7.553 \times 10^{-1}$ & eddy turnover time for the integral scale \\
        $\tau_\eta$ & $1.8367 \times 10^{-6}$ & eddy turnover time for the dissipative scale \\
        $\Delta t$ & $1 \times 10^{-8}$ & timestep of GT \\
        $\Delta \tilde t$ & $1 \times 10^{-5}$ & timestep of LES-NN model \\
        $N_\text{data}$ & $256$ & number of initial conditions of dataset \\
        $N_\text{batch}$ & $1024$ & batch size for training \\
        $T_\text{train}$ & $1.65 \tau_0$ & integration time of training dataset \\
        $T_\text{test}$ & $3.31 \tau_0$ & integration time of test dataset \\
        \bottomrule
    \end{tabular}
    \caption{Values of the parameters of the numerical experiments.}
    \label{tab:params}
\end{table}
\section{Results}
\label{sec:results}

In the following, we present the results from our model and how they compare to the ground truth. In some of the results, we also compare them with state-of-the-art DL closures as well as phenomenological ones.

\begin{figure*}[tb]
    \centering
    \includegraphics[width=.9\linewidth]{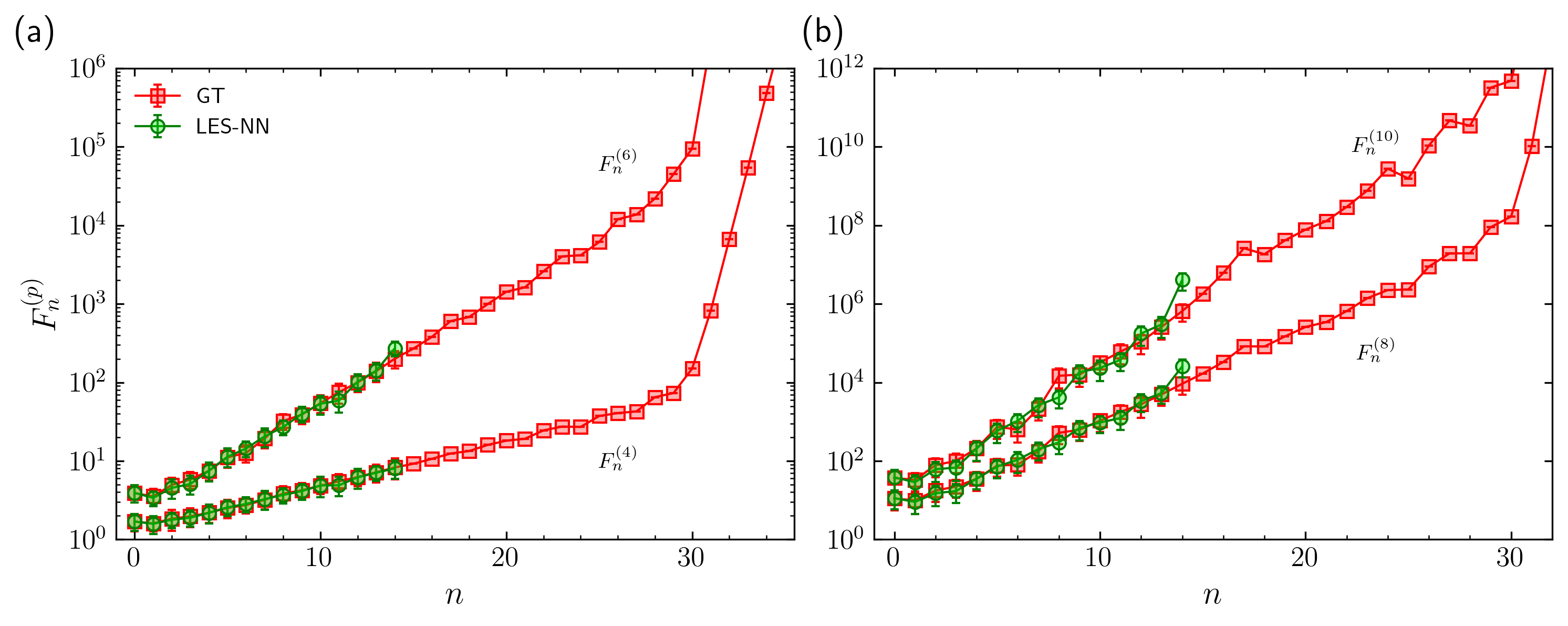}
    \caption{Flatness of different orders computed in terms of the Eulerian structure functions by \autoref{eq:flatness}: (a) flatness of order 4 and 6; (b) flatness of order 8 and 10. Error bars computed by dividing the dataset into chunks, computing the individual chunk's statistics and from here estimate the standard deviation. The error bars are only shown until the cutoff scale.}
    \label{fig:flatness}
\end{figure*}

\autoref{fig:flatness} shows the flatness of different orders, from $F^{(4)}$ and $F^{(10)}$, with respect to the shell index. The flatness is computed in terms of the Eulerian structure functions as:
\begin{equation}
    F^{(p)}_n = \frac{S^{(p)}_n}{(S^{(2)}_n)^{\frac{p}{2}}}\,,
\label{eq:flatness}
\end{equation}
where the Eulerian structure functions are expressed:
\begin{equation}
    S_n^{(p)} = \langle|u_n|^p\rangle_t\,,
\label{eq:esf}
\end{equation}
with $\langle\cdot\rangle$ representing the averaging operator. The lower-order flatnesses show a good agreement with the ground truth. As the order increases, we start to notice some deviations, especially near the cutoff. Despite these deviations, the results remain promising, as these higher-order moments are non-trivial to reproduce correctly, and phenomenological closures fail to capture them accurately.

\begin{figure*}[tb]
    \centering
    \includegraphics[width=.9\linewidth]{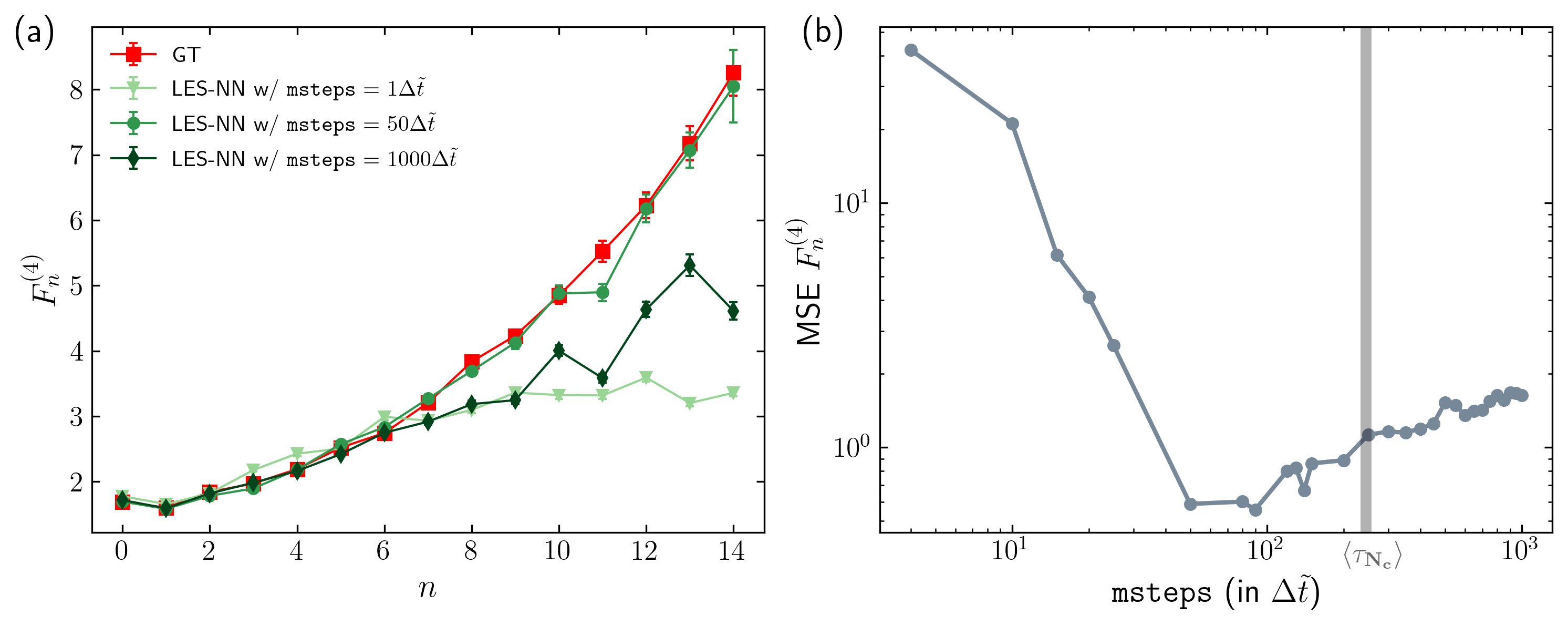}
    \caption{(a) Fourth order flatness for different time in the loop (\texttt{msteps}) and compared with the ground truth (error bars represent the standard deviation). (b) Mean square error of the fourth order flatness (\autoref{eq:mse_f}) for different time in the loop.}
    \label{fig:flatness2}
\end{figure*}

\autoref{fig:flatness2} attempts to determine the optimal time in the loop. On the left, $F_n^{(4)}$ is shown for different times in the loop (the \texttt{msteps} variable used in \autoref{alg:training}) $1 \Delta t$, $50 \Delta t$ and $1000 \Delta t$, where one time step in the loop corresponds to the \emph{a priori} training paradigm. We can see that the best results are obtained with a value of $\texttt{msteps} = 50\Delta t$, while both $\texttt{msteps} = 1\Delta t$ and $\texttt{msteps} = 1000\Delta t$ perform poorly in comparison. In the subfigure on the right, we show a continuation of this analysis, where we plot the MSE of $F_n^{(4)}$, given by 
\begin{equation}
\text{MSE }(F_n^{(4)}) = \frac{\sum_{n = 0}^{N_{c}}|F_{n_{GT}}^{(4)} - F_{n_{LES}}^{(4)}|^2}{N_{c}}\,,
\label{eq:mse_f}
\end{equation}
with respect to the time in the loop. This allows us to better understand the effect of the time in the loop in the effectiveness of the training procedure and how it impacts the final performance of the learned closure. There is a benefit in increasing the time in the loop until around $100 \Delta t$ in the loop. Keeping on increasing after this threshold increases the error. The highest MSE occurs with instantaneous evaluation, i.e., when $\texttt{msteps} = 1$.

We saw that there is a clear effect from the duration in the loop during training in the performance of the model. \emph{A priori}, we expect that the optimal loop time will be a fraction of the eddy turnover time of the fastest shell included in the loss. \newstuff{Since we use a loss function that measures the difference between velocity fields}, exceeding the eddy turnover time of the fastest shell with a high \texttt{msteps} value causes the signals (GT and our model) to decorrelate, making the loss less meaningful. Our focus is on achieving a statistically accurate closure rather than synchronizing with the GT, which is unrealistic. Therefore, the ideal loop time is expected to be a fraction of $\tau_{N_{c}}$. Exceeding this value smooths out the dynamics of the fastest shells, pushing the model to track the moving average of the GT rather than its exact behavior.

\begin{figure*}[tb]
    \centering
    \includegraphics[width=0.9\linewidth]{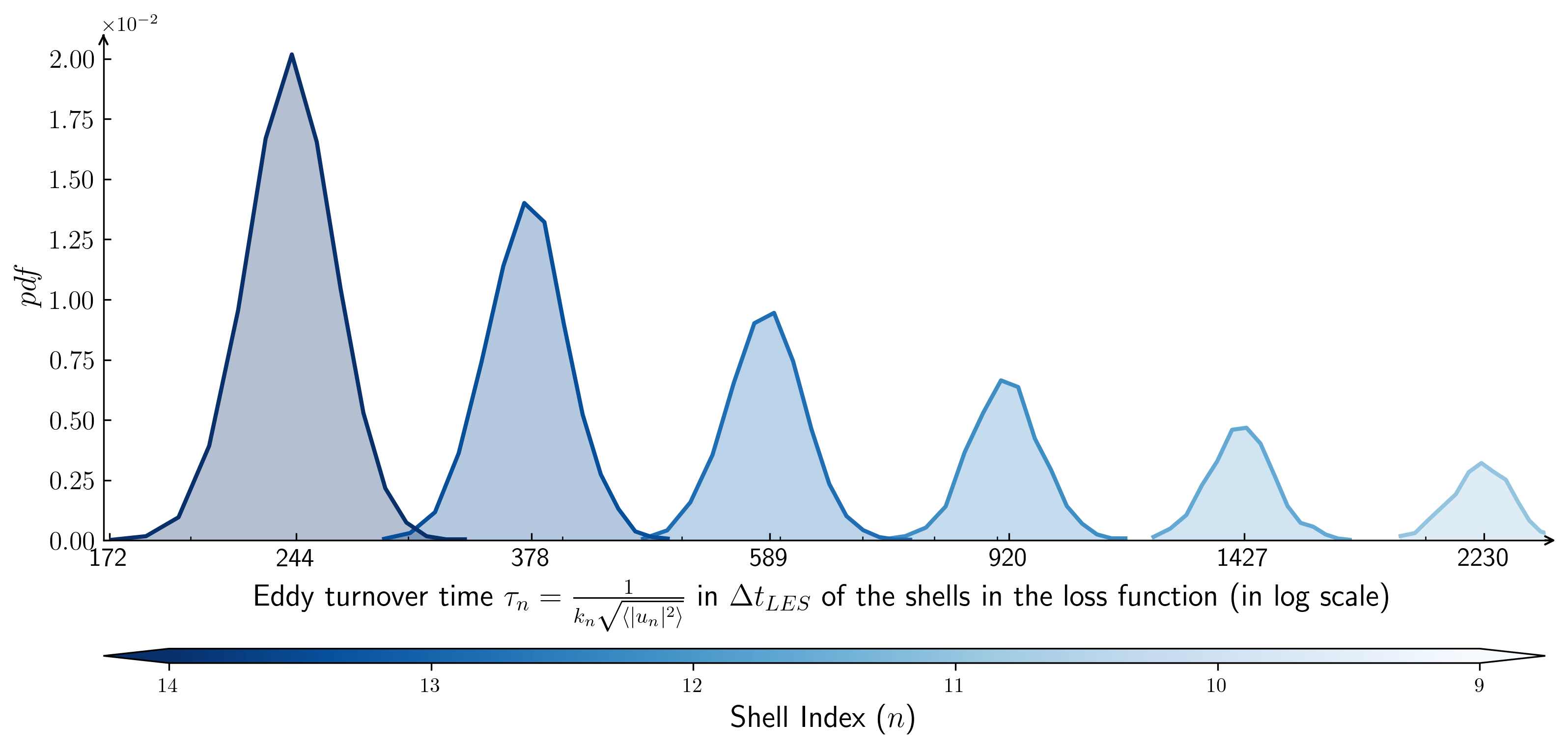}
    \caption{pdf of the eddy turnover time of the shells used in the loss function.}
    \label{fig:time_scales}
\end{figure*}

To better understand this relation between time scales of the system and \texttt{msteps}, we show the probability density function (pdf) of the eddy turnover time $\tau_n$ for the shells considered in the loss in \autoref{fig:time_scales}, computed as 
\begin{equation}
    \tau_n = \frac{1}{k_n \sqrt{\langle|u_n|^2}\rangle_t}\,,
\end{equation}
where the pdf is obtained considering a time signal of $u_n$ for various initial conditions.

Looking back at \autoref{fig:flatness2}, when we examine the MSE of $F^{(4)}_n$, we see that the optimal \texttt{msteps} value corresponds to a fraction of the eddy turnover time of the cutoff shell, $\langle\tau_{N_{c}}\rangle = 244$, with the ideal value being around \texttt{msteps} = 100, or approximately $0.41 \langle\tau_{N_{c}}\rangle$. This analysis shows the benefit of using the \emph{solver-in-the-loop} approach versus the conventional static paradigm and helps understand the physicality of the optimal time in the loop. Throughout the rest of the paper, we will try to keep making similar analyses as we did here for the flatness, for other quantities, as to validate our hypothesis.

\begin{figure}[tb]
    \centering
    \includegraphics[width=\figsize\linewidth]{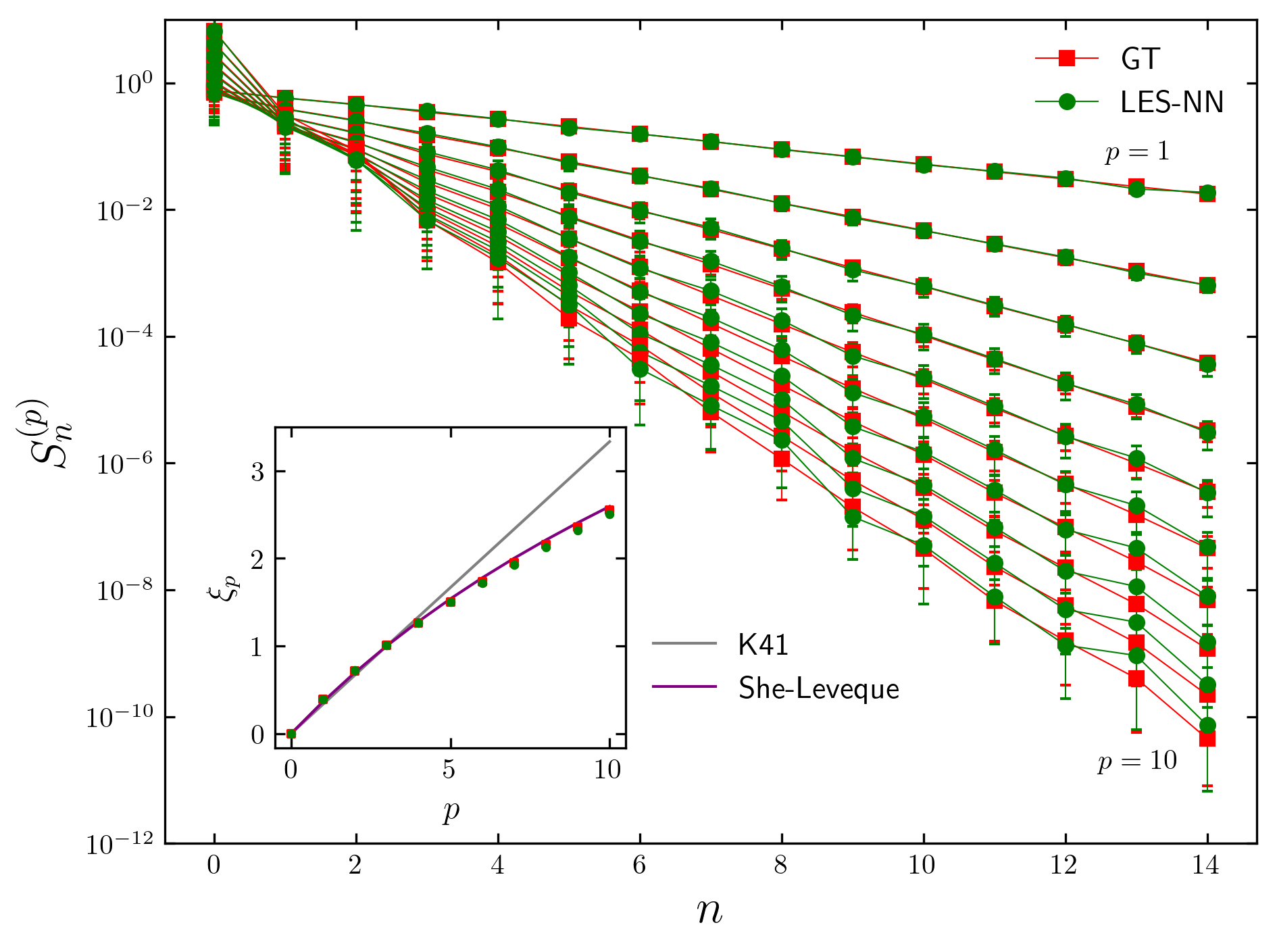}
    \caption{Eulerian structure functions $S_n^{(p)} = \langle|u_n|^p\rangle_t$ vs.
shell index $n$, in lin-log scale, for orders p from 1 to 10
and with $N_{c}$ = 14, comparison between ground truth (GT) and prediction (Pred). Inset plot: Anomalous scaling exponents $\xi_p$ of the Eulerian structure functions $S_n^{(p)} \propto k_n^{-\xi_p}$ for the fully resolved model (GT), our model (LES-NN), the prediction from K41 theory \cite{TikhomirovUnknownTitle1991} and the prediction from She-Leveque model \cite{SheUnknownTitle1994}.}
    \label{fig:esf_asexp}
\end{figure}

\autoref{fig:esf_asexp} shows the Eulerian structure functions. The results from our closure align closely with the GT within error bars, though more noticeable deviations appear as the order increases and near the cutoff. The error-bars are estimated by splitting the datasets in chunks. We compute individual statistics for each chunk, report the average as the central point, and use
the difference between the minimum and maximum as the error bar. To further verify our implementation, we show as an inset plot the anomalous scaling exponents $\xi_p$ of the Eulerian structure functions: 
\begin{equation}
S_n^{(p)} \propto k_n^{-\xi_p}\,.
\label{eq:xi}
\end{equation}
Also here, we see an agreement with the GT similar to what we saw with the flatnesses.

\begin{figure*}[tb]
    \centering
    \includegraphics[width=.9\linewidth]{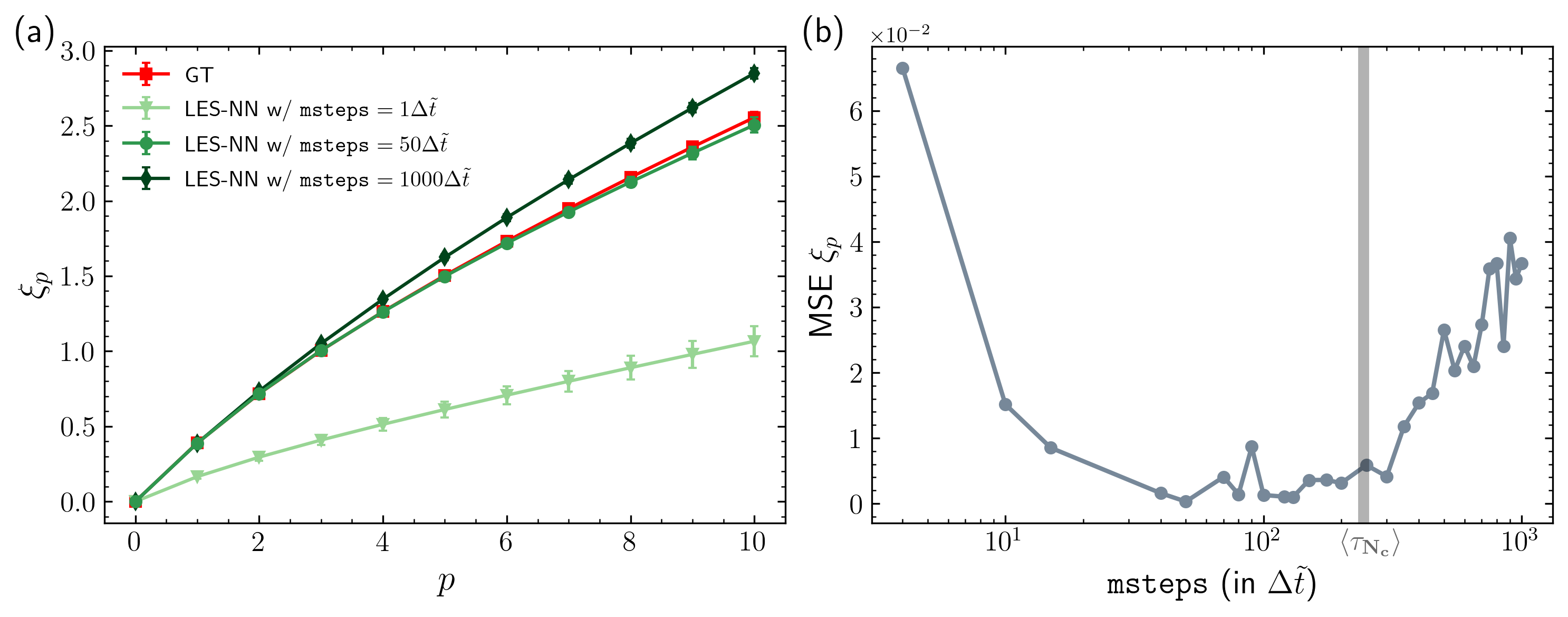}
    \caption{(a) Anomalous scaling exponents $\xi_p$ for different time in the loop (\texttt{msteps}) and compared with the ground truth. (b) MSE of $\xi_p$ (\autoref{eq:mse_asexp}) for different time in the loop.}
    \label{fig:asexp_comp}
\end{figure*}

Looking deeper into the anomalous scaling exponents, \autoref{fig:asexp_comp} shows on the left the comparison of this quantity for different time in the loop and on the right the MSE computed via \autoref{eq:mse_asexp}. On the left, we see similar results as we saw before, where a value of \texttt{msteps} $= 50 \Delta t$ performs best. 
\begin{equation}
\text{MSE }(\xi_p) = \frac{\sum_{p = 1}^{P = 10}|\xi_{p_{GT}} - \xi_{p_{LES}}|^2}{P}
\label{eq:mse_asexp}
\end{equation}
The MSE of the anomalous scaling exponents is even more expressive than the one of the flatness, as it incorporates statistical moments from $p = 1$ to $p = 10$ (and since it is not normalised, it gives more weight to higher order ones). Similar to the flatness case, the ideal loop time is a fraction of the eddy turnover time of the fastest shell. Deviating too much from this value, either higher or lower, results in an increase in MSE.

\begin{figure}[tb]
    \centering
    \includegraphics[width=\figsize\linewidth]{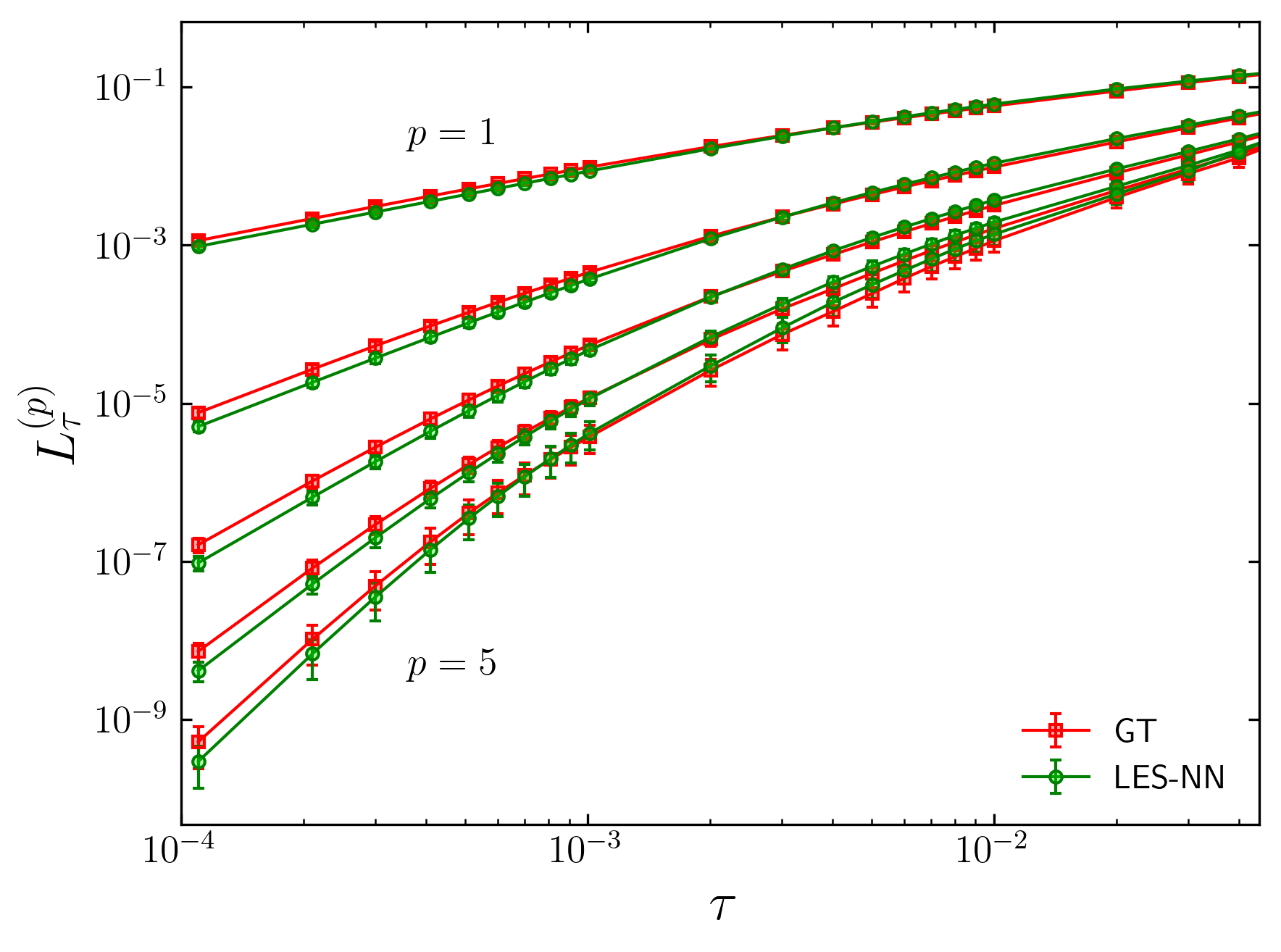}
    \caption{Lagrangian structure functions or orders $p = 1, \ldots, 5$ in log-log scale with the time lag, $\tau$, on the x-axis.}
    \label{fig:lsf}
\end{figure}

Shifting the perspective from Eulerian to Lagrangian, we now examine the Lagrangian structure functions, assessing whether our model accurately reproduces time correlations across various time lags. This is illustrated in \autoref{fig:lsf}. 

The Lagrangian structure functions are computed as
\begin{equation}
    L_{\tau}^{(p)} = \langle|u(t + \tau) - u(t)|^p\rangle_t\,,
\end{equation}
where the Lagrangian signals are obtained by summing the real parts of all the shells $u(t) = \Re(\sum_n u_n(y))$. Analysing the results, one can see that the model closely follows the scaling of the ground truth, even for small time lags and higher-order moments (within error bars), which are the most challenging to capture accurately.

\begin{figure*}[tb]
    \centering
    \includegraphics[width=1.0\linewidth]{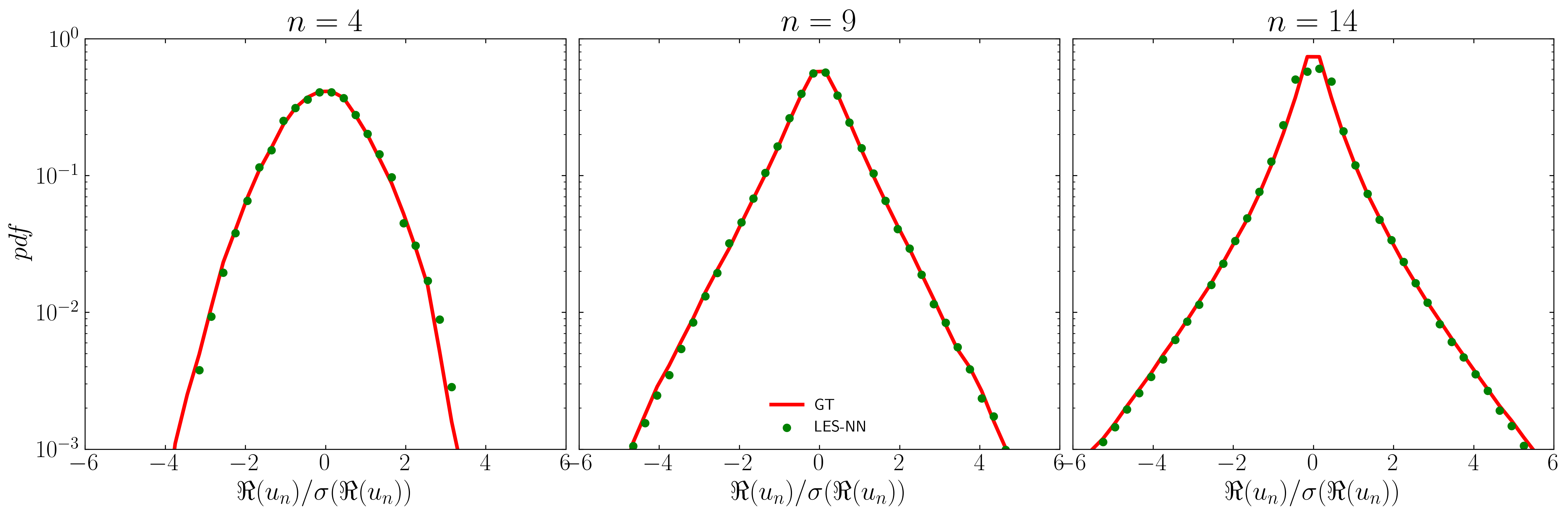}
    \caption{pdf of the real part of the shells 4, 9 and 14 (cutoff shell), in log scale, normalized with the standard deviation $\Re(u_n) / \sigma (\Re(u_n))$ for the GT and prediction.}
    \label{fig:pdf_re_u}
\end{figure*}

In \autoref{fig:pdf_re_u}, we show another statistical quantity: the pdf of the real part of the velocity signals for different shells $n = 4, 9, 14$ normalized by the standard deviation, for both the model and the ground truth. We see that our closure has the correct effect on the resolved scales as we are able to correctly reproduce the Gaussian statistics of the large scales and more importantly the non-Gaussian statistics of the small scales, characterized by intermittency.

\begin{figure*}[tb]
    \centering
    \includegraphics[width=1.0\linewidth]{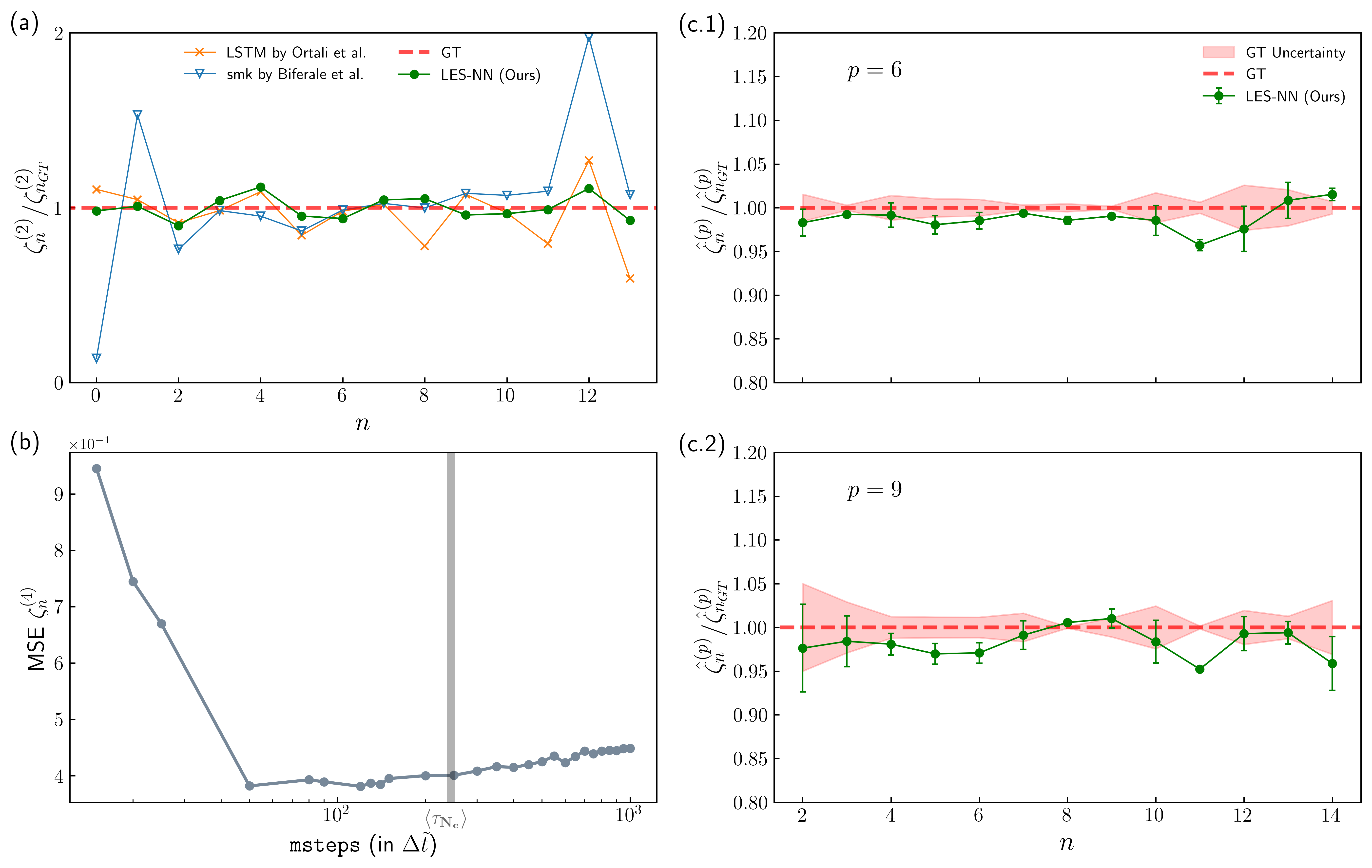}
    \caption{(a) The local slopes (normalised by the ones of the GT) for the second order Eulerian structure functions (\autoref{eq:slopes}) vs. shell index $n$. Comparison between ground truth (GT), our model (LES-NN), the state-of-the-art LSTM approach by Ortali et al. \cite{OrtaliUnknownTitle2022} and a non-ML approach of an optimal subgrid closure scheme by Biferale et al. \cite{BiferaleUnknownTitle2017}. (b) MSE of $\zeta_p^{(4)}$ (\autoref{eq:mse_zeta}) for different time in the loop. (c) Normalised local slopes of the Eulerian structure functions computed using the triads (\autoref{eq:s_pi}) for $p = 6$ and $p = 9$. Remark: this expression to compute the Eulerian structure functions using the triads, unlike $\langle|u_n|^p\rangle$, is not prone to period-3 oscillations and as such the local slopes oscillate much less. This is why even for such high-order moments slopes as the ones in (c.1) and (c.2) the  range of values is much closer to the GT than in (a).}
    \label{fig:slopes}
\end{figure*}

We aim also to compare our closure with other state-of-the-art closures. As such, in \autoref{fig:slopes}-(a), we show the local slopes for the second-order Eulerian structure function, expressed through
\begin{equation}
\zeta_n^{(p)}=\frac{\log \left[S_{n+1}^{(p)}\right]-\log \left[S_n^{(p)}\right]}{\log [\lambda]}\,,
\label{eq:slopes}
\end{equation}
with respect to the shell index, for our model, the GT, the Long Short-Term Memory (LSTM) \cite{hochreiter1997lstm} approach from Ortali et al. \cite{OrtaliUnknownTitle2022} and a phenomenological closure from Biferale et al. \cite{BiferaleUnknownTitle2017}. The LSTM approach performs well overall although its accuracy decreases near the cutoff. Its memory component compensates for the static training, contributing to its robust performance. The phenomenological closure, \emph{smk}, performs adequately in the mid-range but shows significant degradation near the boundaries. In contrast, our approach oscillates around the GT and achieves the best performance of the three models, particularly near the cutoff — where correctly reproducing the slopes is most challenging due to the more pronounced effect of the closure.

\autoref{fig:slopes}.b shows the MSE of the local slopes of the fourth order Eulerian structure function:
\begin{equation}
\text{MSE }(\zeta_n^{(4)}) = \frac{\sum_{n = 0}^{N_{c}}|\zeta_{n_{GT}}^{(4)} - \zeta_{n_{LES}}^{(4)}|^2}{N_{c} - 1}\,.
\label{eq:mse_zeta}
\end{equation}
The trend is similar to the one seen before, although here the increase in error with high \texttt{msteps} values is not so severe as before. The way to interpret this trend is that for high \texttt{msteps}, the slope of $S_n{\!}^{(4)}$ is correct, but the total energy content is off (see the vertical shift of the structure functions). The inter-shell relations are preserved, but the absolute energy is inaccurate. This also explains why the anomalous scaling exponents showed the most difference out of the three MSE errors for high \texttt{msteps}: it considers very high order moments and gives them a considerable weight.

Lastly, in \autoref{fig:slopes}-(c), we show the normalised local slopes of the Eulerian structure functions computed with respect to the triads. These structure functions are computed from \autoref{eq:s_pi}. Unlike the ones from \autoref{eq:esf}, these are not prone to period-3 oscillations. The slopes are computed using the same expression as before, \autoref{eq:slopes}. As expected, performance degrades as the cut-off is approached. Despite this, for such a sensitive quantity as the local slopes of structure functions, our model remains relatively close to the ground truth, with errors of less than 5\%. 
\begin{equation}
    \hat{S}_n^{(p)} = \langle|u^*_{n-2} u^*_{n-1} u_{n}|^{\frac{p}{3}}\rangle_t
\label{eq:s_pi}
\end{equation}

\begin{figure}[tb]
    \centering
    \includegraphics[width=\figsize\linewidth]{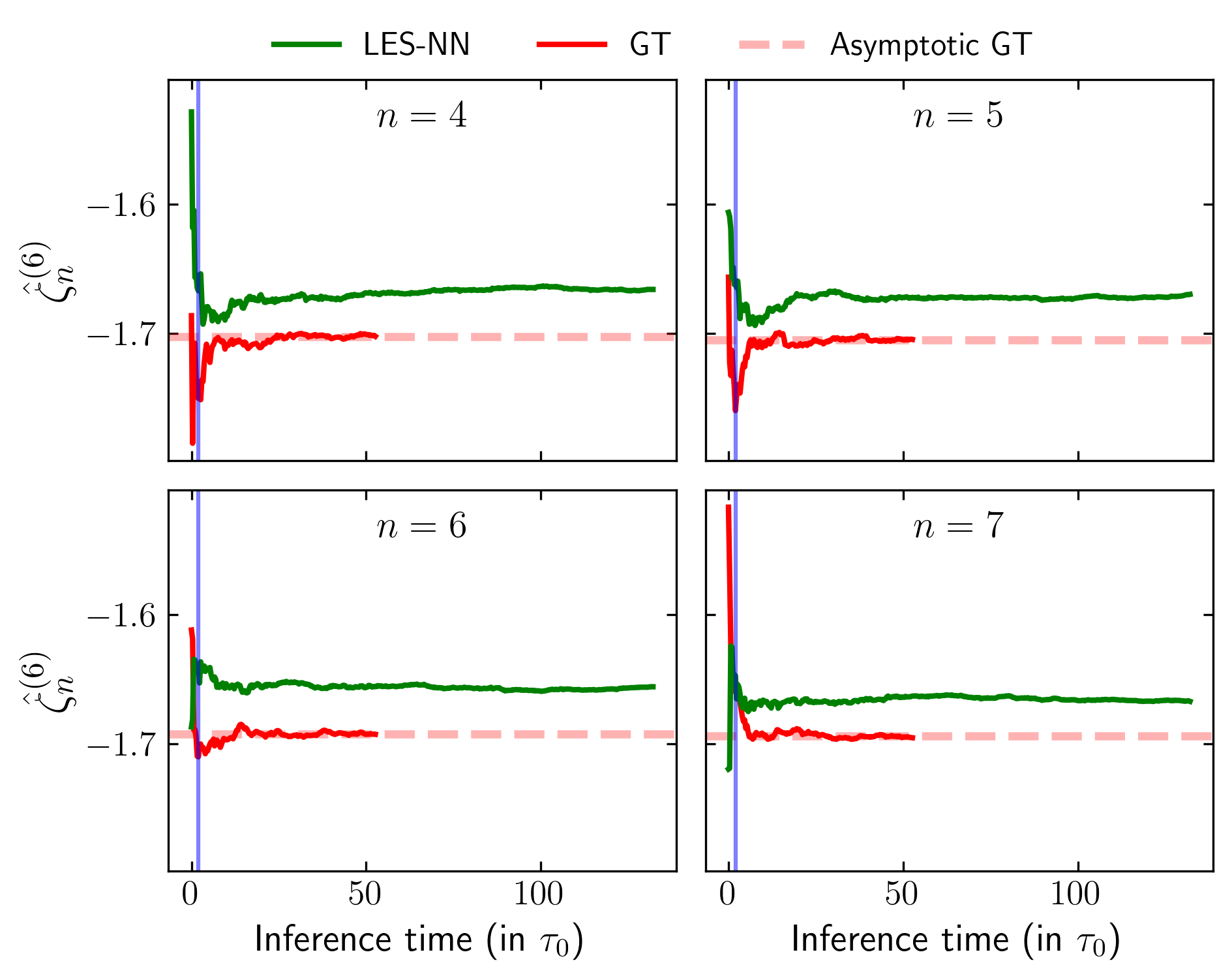}
    \caption{Local slopes of the sixth order Eulerian structure function computed as a function of the triads (\autoref{eq:s_pi}), $\hat{\zeta}^{(6)}_n$, with respect to the inference/deployment time (expressed in terms of $\tau_0$). It is shown for $n = 4,5,6,7$. Compared with the GT reference data. The vertical bar denotes the amount of data used for training data.}
    \label{fig:inference_time}
\end{figure}

As demonstrated in previous figures, our model exhibits some error relative to the ground truth. It is important to determine whether this error arises solely from the model's inherent limitations or if a significant statistical error is also present, potentially due to computing a given statistical observable from a limited sample size. \autoref{fig:inference_time} explores this issue by showing how the local slope of the third-order Eulerian structure function computed using the triads (\autoref{eq:s_pi}) evolves over increasing deployment time, for shells $n = 4, 5, 6, \text{ and } 7$. The figure presents results for both our model (LES-NN) and the ground truth.

We observe that only a few eddy turnover times are needed to approach the asymptotic value, both for the ground truth and our model. This suggests that the errors highlighted throughout the paper are primarily due to intrinsic model limitations rather than statistical fluctuations (with the obvious caveat that error bars, when shown, refer to statistical errors). The vertical bar in the figure represents the amount of data used to train our model. Notably, the model remains stable even when deployed far beyond the time frame it was trained on. This stability naturally arises from our training methodology, where we explicitly constrain the time evolution based on the actual governing equations.

The GT is shown for less deployment time than the LES-NN model because it takes much longer to run. We benchmarked the time it takes to run them both on an NVIDIA A100 GPU, averaging over many realizations to ensure the validity of our results. We started with an ensemble of initial conditions and evolved them for one eddy turnover time of the slowest shell, $\tau_0$. On average, the GT took about 81 minutes, whereas the LES-NN only took 6 minutes.

This difference can be attributed to the higher time step used in the LES. Although the presence of the neural network introduces some computational overhead, our fully differentiable framework allows us to accelerate computations by utilizing graph mode and XLA (Accelerated Linear Algebra) compilation \cite{xla}. Graph mode enables numerous optimizations at the compiler level, such as statically determining the values of tensors by combining constant nodes in the computation, commonly referred to as ``constant folding.'' XLA allows for the optimization of the computational graph. One such optimization is for example the separation of independent parts of a computation, enabling them to be processed across multiple threads or devices. This parallelism enhances performance significantly. Furthermore, XLA simplifies arithmetic operations by eliminating common subexpressions, leading to a more efficient execution of the model.

\begin{figure}[tb]
    \centering
    \includegraphics[width=\figsize\linewidth]{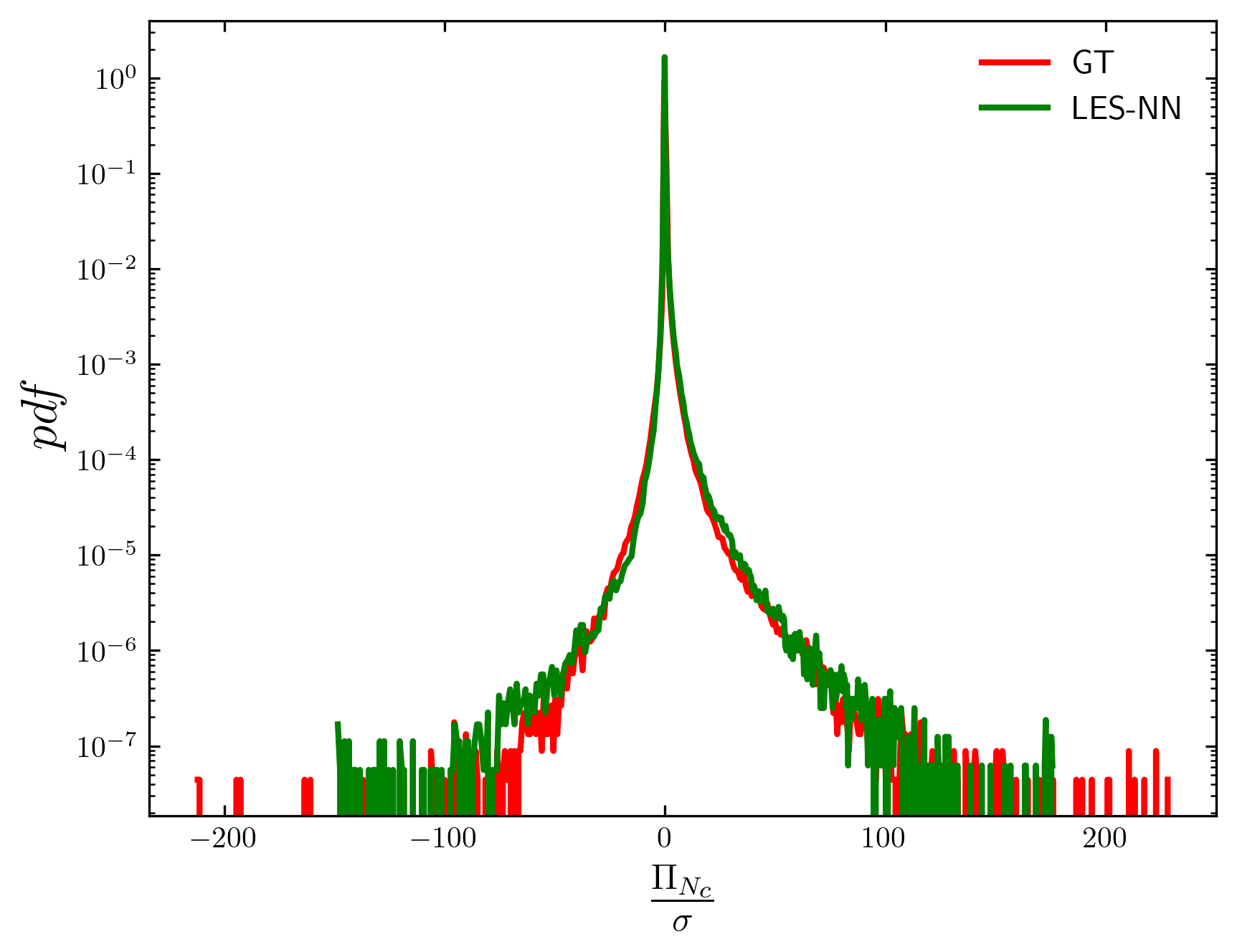}
    \caption{pdf of the flux at shell $N_{c}$ (\autoref{eq:flux}).}
    \label{fig:fluxes}
\end{figure}

To evaluate the correct reproduction of the energy fluxes given our closure, we show the pdf of the convective fluxes at the cutoff shell, $\Pi_{N_{c}}$ in \autoref{fig:fluxes}. Where $\Pi_n$ is given by: 
\begin{equation} 
\Pi_n= \Im[ak_{n+1}u_{n+2}u_{n+1}^*u_n^* + (b+a)k_nu_{n+1}u_n^*u_{n-1}^*].
\label{eq:flux}
\end{equation}
The results show a strong agreement with the fully resolved model. A positive value of this flux indicates a forward energy cascade at the cutoff shell, transferring energy to smaller scales. Conversely, a negative value is called backscatter, meaning energy flows from smaller scales to larger ones. This phenomenon is particularly challenging to model in subgrid-scale models, as improper handling of negative energy flux can lead to numerical instabilities. This is why phenomenological closures often avoid addressing backscatter. Similarly, some deep learning-based closures sidestep this issue to simplify training and ensure model stability.

\begin{figure}[tb]
    \centering
    \includegraphics[width=\figsize\linewidth]{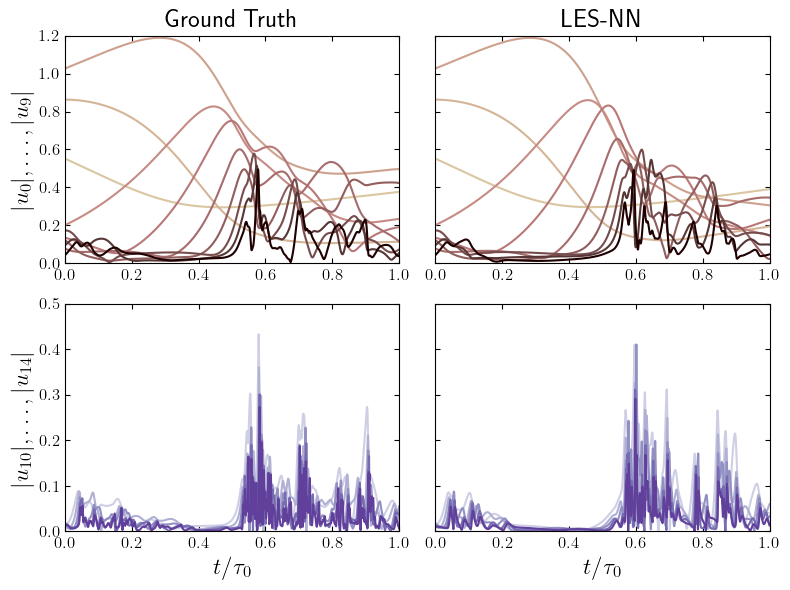
    }
    \caption{Qualitative comparison between the GT and Prediction in terms of the dynamics of the absolute value of the large scales, $|u_0|, \ldots, |u_9|$ and the small scales, $|u_{10}|, \ldots, |u_{14}|$.}
    \label{fig:signals}
\end{figure}

\autoref{fig:signals} shows a comparison of simulation results over a selected time interval. The top row depicts the large scales, where the dynamics between the ground truth and our model remain qualitatively similar until about $t = 0.5 \tau_0$. After this point, the phases begin to decorrelate, especially for the smallest large-scale components (shown in darker colors). The small scales (bottom row) become fully decorrelated at around the same time, but it occurs more rapidly due to their shorter eddy turnover times. 

It is unreasonable to expect a subgrid-scale model to maintain synchronization between the LES model and the GT for extended periods. The goal is simply to recover the statistical moments of the GT rather than precisely replicate its dynamic evolution.

\begin{figure}[tb]
    \centering
    \includegraphics[width=\figsize\linewidth]{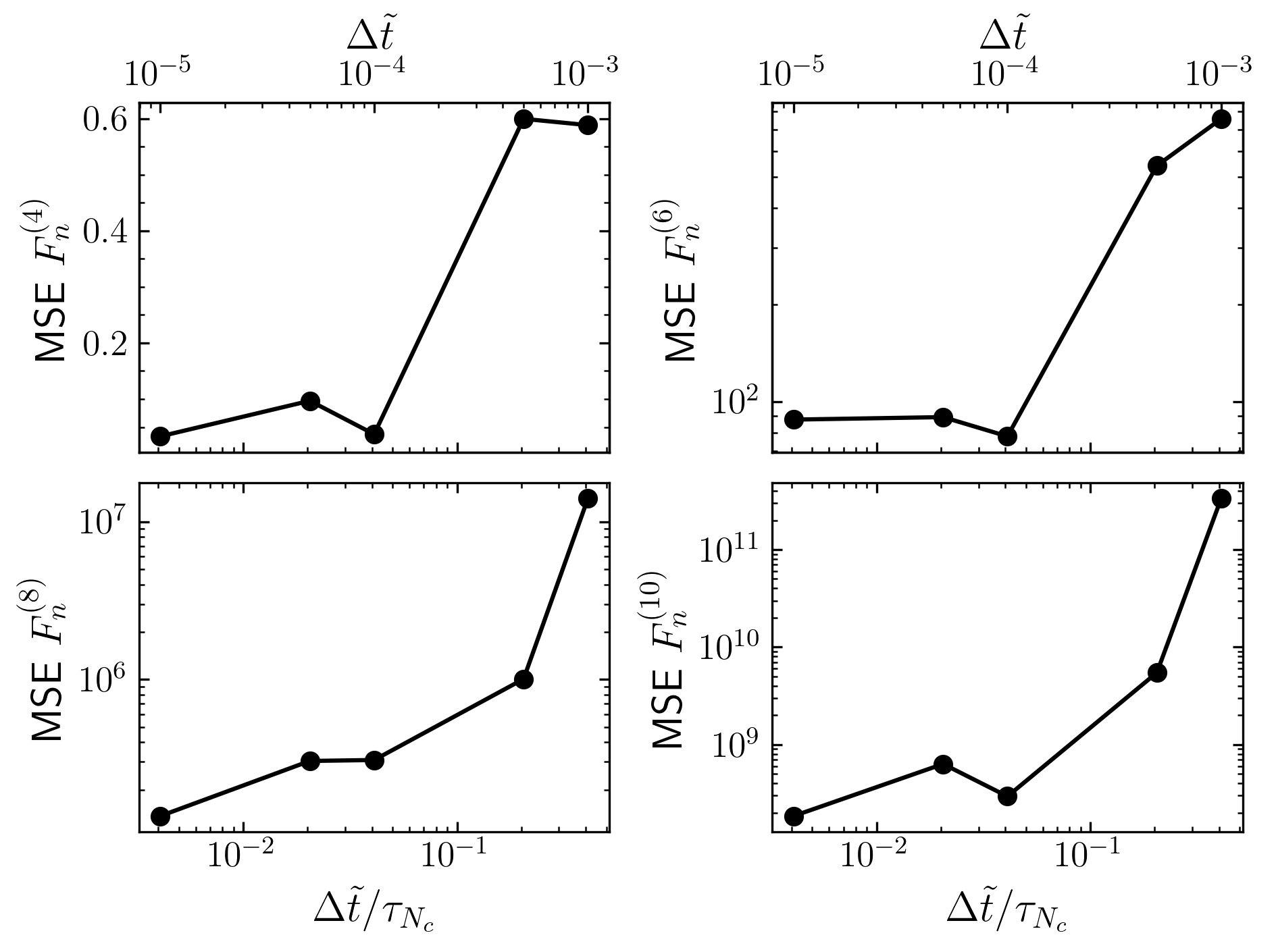}
    \caption{Mean square error for the flatness of different orders, $F_n^{(4)}$, \ldots, $F_n^{(10)}$, for LES-NN models trained with different time steps, $\Delta \tilde t$. The error is computed with respect to the ground truth. The time step is normalised by the time scale of the cutoff shell.}
    \label{fig:time_step_study}
\end{figure}

The time step used for the LES, $\Delta \tilde{t} = 10^{-5}$, was chosen simply because it was the one used by Ortali et al \cite{OrtaliUnknownTitle2022}. Given that our model performs best when trained for an unrolled time of approximately 250 LES time steps ($\approx\langle\tau_{N_c}\rangle$) or fewer, we were curious to see how the model would behave if we increased the time step up to the limit where only two steps are performed in the loop (\texttt{msteps} $= 2\Delta \tilde t$), with a time step of $10^{-3}$. This analysis is presented in \autoref{fig:time_step_study}, where we plot the MSE of the flatness for various orders, ranging from 4 to 10, as a function of the LES time step.

Surprisingly, we observe that the model maintains very good performance even with a time step 10 times larger than what was used throughout the paper. It is important to note that a time step of $\Delta \tilde{t} = 10^{-4}$ is 10,000 times larger than the ground truth evolution's time step. This indicates that the neural network is not only learning the physical closure but also some numerical error associated with such coarse temporal dynamics. However, when the time step is further increased to $\Delta t = 5\times10^{-4}$ or $\Delta t = 10^{-3}$, the errors grow significantly, and a noticeable drop in performance occurs. This is due to two factors: the increasing influence of numerical errors, which makes the task more challenging for the neural network, and the reduced number of unrolled steps during training, as larger time steps reach the limit of $\tau_{N_c}$ more quickly.

\section{Conclusions}
\label{sec:conclusion}

In this work, we have proposed a solver-in-the-loop approach to learning subgrid-scale closures in a shell model of turbulence. This methodology leverages the differentiable physics paradigm, allowing the neural network to interact with the solver during training and optimize the closure terms a posteriori. By incorporating unrolled solver interactions, we have demonstrated that our model outperforms traditional a-priori trained models in terms of stability and accuracy. Moreover, we show that our model is able to perform similarly or even outperform state-of-the-art deep learning approaches with complex architectures, despite relying on a simpler architecture.


\newstuff{Our results on shell models suggest that the optimal time-in-the-loop is closely tied to the eddy turnover time of the fastest shells included in the loss function. This time scale serves as an approximation of the Lyapunov time of the LES system.
When using a loss function based on the difference between velocities, setting the time-in-the-loop beyond the Lyapunov time effectively attempts to synchronize two systems beyond their synchronization time. Moreover, as the time-in-the-loop increases, the gradients during backpropagation become more likely to either explode or vanish, a phenomenon related to the Lyapunov time, making gradient-based optimization increasingly unstable.
We speculate that even if a different loss function was used, e.g. one based on difference in energy flux between the ground truth and the model, instead of difference in velocities, the optimal time-in-the-loop would not differ significantly from the one identified in our study. While such a loss function does not explicitly enforce trajectory synchronization and is therefore not directly constrained by the Lyapunov time, the gradients’ quality deteriorates as the time-in-the-loop exceeds the Lyapunov time, which might lead to divergence in training or “worse” optimizer steps.}

\newstuff{
Extending these conclusions to the 3D NS introduces additional challenges, primarily due to the increased memory requirements. The general principles derived from our study may still hold, though this remains to be investigated.
}

\newstuff{Our study also provides relevant insights for future work using the solver-in-the-loop approach, particularly in how to \emph{a priori} tune this time-in-the-loop hyperparameter. These insights are grounded in the physics of the system, allowing for better generalization across different physical models.
}

Beyond shell models, we believe this approach shows great potential for more complex systems, including Navier-Stokes equations, where unresolved scales play a much more complex role, making learning closure models more challenging. As such, future work includes extending this framework to NSE turbulence. \newstuff{A potential candidate is natural convection in 2D, where the presence of non-trivial multifractal scaling properties for temperature and Bolgiano scaling for velocity make the closure problem potentailly as challenging as in 3D turbulence, still retaining a smaller degree of complexity.} Additionally, the use of differentiable solvers opens up new possibilities for integrating physical priors more deeply into machine learning frameworks, potentially further improving generalization and data efficiency. This insight is not limited to subgrid-scale closure in the context of LES but can also benefit other problems in fluid dynamics and more generally in science where machine learning can provide solutions.

The solver-in-the-loop methodology can also be compared to reinforcement learning (RL) approaches, which similarly aim to optimize decision making through iterative feedback. While model-free RL typically involves exploring a vast action space and learning from trial and error, our approach directly integrates the physics of the problem, leveraging the differentiable nature of the solver to guide the neural network training. However, when considering model-based RL, the distinctions between our solver-in-the-loop approach and RL become less clear. Model-based RL uses an explicit model of the environment to predict future states and optimize actions, similar to how our methodology utilizes a differentiable physics model during training. The solver in the loop approach is conceptually similar to model-based deep RL, i.e. when the policy is parameterized by a (deep) neural network. This overlap raises interesting questions about the advantages and disadvantages of each approach.

In conclusion, the solver-in-the-loop approach presents a robust and flexible method for addressing subgrid-scale modeling challenges in turbulence using deep learning. We believe it provides a valuable perspective for combining machine learning with differentiable physics to tackle complex, multiscale systems.

\section*{Acknowledgements}
\raggedbottom
The authors benefited from discussions with M. Sbragaglia. This research was supported by European Union’s HORIZON MSCA Doctoral Networks programme under Grant Agreement No. 101072344, project AQTIVATE (Advanced computing, QuanTum algorIthms and data-driVen Approaches for science, Technology and Engineering), the European Research Council (ERC) under the European Union’s Horizon 2020 research and innovation programme Smart-TURB (Grant Agreement No. 882340), and through an Inria Chair.

\pagebreak

\bibliography{apssamp}

\begin{thebibliography}{43}%
\makeatletter
\providecommand \@ifxundefined [1]{%
 \@ifx{#1\undefined}
}%
\providecommand \@ifnum [1]{%
 \ifnum #1\expandafter \@firstoftwo
 \else \expandafter \@secondoftwo
 \fi
}%
\providecommand \@ifx [1]{%
 \ifx #1\expandafter \@firstoftwo
 \else \expandafter \@secondoftwo
 \fi
}%
\providecommand \natexlab [1]{#1}%
\providecommand \enquote  [1]{``#1''}%
\providecommand \bibnamefont  [1]{#1}%
\providecommand \bibfnamefont [1]{#1}%
\providecommand \citenamefont [1]{#1}%
\providecommand \href@noop [0]{\@secondoftwo}%
\providecommand \href [0]{\begingroup \@sanitize@url \@href}%
\providecommand \@href[1]{\@@startlink{#1}\@@href}%
\providecommand \@@href[1]{\endgroup#1\@@endlink}%
\providecommand \@sanitize@url [0]{\catcode `\\12\catcode `\$12\catcode `\&12\catcode `\#12\catcode `\^12\catcode `\_12\catcode `\%12\relax}%
\providecommand \@@startlink[1]{}%
\providecommand \@@endlink[0]{}%
\providecommand \url  [0]{\begingroup\@sanitize@url \@url }%
\providecommand \@url [1]{\endgroup\@href {#1}{\urlprefix }}%
\providecommand \urlprefix  [0]{URL }%
\providecommand \Eprint [0]{\href }%
\providecommand \doibase [0]{https://doi.org/}%
\providecommand \selectlanguage [0]{\@gobble}%
\providecommand \bibinfo  [0]{\@secondoftwo}%
\providecommand \bibfield  [0]{\@secondoftwo}%
\providecommand \translation [1]{[#1]}%
\providecommand \BibitemOpen [0]{}%
\providecommand \bibitemStop [0]{}%
\providecommand \bibitemNoStop [0]{.\EOS\space}%
\providecommand \EOS [0]{\spacefactor3000\relax}%
\providecommand \BibitemShut  [1]{\csname bibitem#1\endcsname}%
\let\auto@bib@innerbib\@empty
\bibitem [{\citenamefont {Frisch}(1995)}]{frisch}%
  \BibitemOpen
  \bibfield  {author} {\bibinfo {author} {\bibfnamefont {U.}~\bibnamefont {Frisch}},\ }\bibfield  {title} {\bibinfo {title} {Turbulence: The legacy of {A.} {N.} {K}olmogorov},\ }\bibfield  {journal} {\bibinfo  {journal} {Cambride University press}\ }\href {https://doi.org/10.1017/CBO9781139170666} {10.1017/CBO9781139170666} (\bibinfo {year} {1995})\BibitemShut {NoStop}%
\bibitem [{\citenamefont {Alexakis}\ and\ \citenamefont {Biferale}(2018)}]{cascades}%
  \BibitemOpen
  \bibfield  {author} {\bibinfo {author} {\bibfnamefont {A.}~\bibnamefont {Alexakis}}\ and\ \bibinfo {author} {\bibfnamefont {L.}~\bibnamefont {Biferale}},\ }\bibfield  {title} {\bibinfo {title} {Cascades and transitions in turbulent flows},\ }\bibfield  {journal} {\bibinfo  {journal} {Physics Reports}\ }\href {https://doi.org/10.1016/j.physrep.2018.08.001} {10.1016/j.physrep.2018.08.001} (\bibinfo {year} {2018})\BibitemShut {NoStop}%
\bibitem [{\citenamefont {Meneveau}\ and\ \citenamefont {Katz}(2000)}]{les_review}%
  \BibitemOpen
  \bibfield  {author} {\bibinfo {author} {\bibfnamefont {C.}~\bibnamefont {Meneveau}}\ and\ \bibinfo {author} {\bibfnamefont {J.}~\bibnamefont {Katz}},\ }\bibfield  {title} {\bibinfo {title} {Scale-invariance and turbulence models for large-eddy simulation},\ }\bibfield  {journal} {\bibinfo  {journal} {Annual Review Fluid Mechanics}\ }\href {https://doi.org/10.1146/annurev.fluid.32.1.1} {10.1146/annurev.fluid.32.1.1} (\bibinfo {year} {2000})\BibitemShut {NoStop}%
\bibitem [{\citenamefont {M.~Lesieur}(2005)}]{lesieur2005}%
  \BibitemOpen
  \bibfield  {author} {\bibinfo {author} {\bibfnamefont {P.~C.}\ \bibnamefont {M.~Lesieur}, \bibfnamefont {O.~Metáis}},\ }\href {https://doi.org/https://doi.org/10.1017/CBO9780511755507} {\emph {\bibinfo {title} {Large-{E}ddy {S}imulations of {T}urbulence}}}\ (\bibinfo  {publisher} {Cambridge Unversity Press},\ \bibinfo {year} {2005})\BibitemShut {NoStop}%
\bibitem [{\citenamefont {Pope}(2001)}]{pope2001}%
  \BibitemOpen
  \bibfield  {author} {\bibinfo {author} {\bibfnamefont {S.~B.}\ \bibnamefont {Pope}},\ }\href {https://doi.org/https://doi.org/10.1017/CBO9780511840531} {\emph {\bibinfo {title} {Turbulent Flows}}}\ (\bibinfo  {publisher} {IOP Publishing},\ \bibinfo {year} {2001})\BibitemShut {NoStop}%
\bibitem [{\citenamefont {Sagaut}(2006)}]{sagaut2006}%
  \BibitemOpen
  \bibfield  {author} {\bibinfo {author} {\bibfnamefont {P.}~\bibnamefont {Sagaut}},\ }\href {https://doi.org/https://doi.org/10.1007/b137536} {\emph {\bibinfo {title} {Large Eddy Simulation for Incompressible Flows: An Introduction}}}\ (\bibinfo  {publisher} {Springer},\ \bibinfo {year} {2006})\BibitemShut {NoStop}%
\bibitem [{\citenamefont {Mailybaev}(2021)}]{AAM21}%
  \BibitemOpen
  \bibfield  {author} {\bibinfo {author} {\bibfnamefont {A.~A.}\ \bibnamefont {Mailybaev}},\ }\bibfield  {title} {\bibinfo {title} {Hidden scale invariance of intermittent turbulence in a shell model},\ }\href {https://doi.org/10.1103/PhysRevFluids.6.L012601} {\bibfield  {journal} {\bibinfo  {journal} {Phys. Rev. Fluids}\ }\textbf {\bibinfo {volume} {6}},\ \bibinfo {pages} {L012601} (\bibinfo {year} {2021})}\BibitemShut {NoStop}%
\bibitem [{\citenamefont {Mailybaev}\ and\ \citenamefont {Thalabard}(2022)}]{Mailybaev2022}%
  \BibitemOpen
  \bibfield  {author} {\bibinfo {author} {\bibfnamefont {A.~A.}\ \bibnamefont {Mailybaev}}\ and\ \bibinfo {author} {\bibfnamefont {S.}~\bibnamefont {Thalabard}},\ }\bibfield  {title} {\bibinfo {title} {Hidden scale invariance in navier–stokes intermittency},\ }\href {https://doi.org/10.1098/rsta.2021.0098} {\bibfield  {journal} {\bibinfo  {journal} {Philosophical Transactions of the Royal Society A: Mathematical, Physical and Engineering Sciences}\ }\textbf {\bibinfo {volume} {380}},\ \bibinfo {pages} {20210098} (\bibinfo {year} {2022})}\BibitemShut {NoStop}%
\bibitem [{\citenamefont {Maulik}\ \emph {et~al.}(2019{\natexlab{a}})\citenamefont {Maulik}, \citenamefont {San}, \citenamefont {Rasheed},\ and\ \citenamefont {Vedula}}]{rm18}%
  \BibitemOpen
  \bibfield  {author} {\bibinfo {author} {\bibfnamefont {R.}~\bibnamefont {Maulik}}, \bibinfo {author} {\bibfnamefont {O.}~\bibnamefont {San}}, \bibinfo {author} {\bibfnamefont {A.}~\bibnamefont {Rasheed}},\ and\ \bibinfo {author} {\bibfnamefont {P.}~\bibnamefont {Vedula}},\ }\bibfield  {title} {\bibinfo {title} {Sub-grid modelling for two-dimensional turbulence using neural networks},\ }\bibfield  {journal} {\bibinfo  {journal} {Journal of Fluid Mechanics}\ }\href {https://doi.org/10.1017/jfm.2018.770} {10.1017/jfm.2018.770} (\bibinfo {year} {2019}{\natexlab{a}})\BibitemShut {NoStop}%
\bibitem [{\citenamefont {Maulik}\ \emph {et~al.}(2019{\natexlab{b}})\citenamefont {Maulik}, \citenamefont {San}, \citenamefont {Jacob},\ and\ \citenamefont {Crick}}]{rm19}%
  \BibitemOpen
  \bibfield  {author} {\bibinfo {author} {\bibfnamefont {R.}~\bibnamefont {Maulik}}, \bibinfo {author} {\bibfnamefont {O.}~\bibnamefont {San}}, \bibinfo {author} {\bibfnamefont {J.~D.}\ \bibnamefont {Jacob}},\ and\ \bibinfo {author} {\bibfnamefont {C.}~\bibnamefont {Crick}},\ }\bibfield  {title} {\bibinfo {title} {Sub-grid scale model classification and blending through deep learning},\ }\bibfield  {journal} {\bibinfo  {journal} {Journal of fluid mechanics}\ }\href {https://doi.org/https://doi.org/10.1017/jfm.2019.254} {https://doi.org/10.1017/jfm.2019.254} (\bibinfo {year} {2019}{\natexlab{b}})\BibitemShut {NoStop}%
\bibitem [{\citenamefont {Beck}\ \emph {et~al.}(2019)\citenamefont {Beck}, \citenamefont {Flad},\ and\ \citenamefont {Munz}}]{beck1}%
  \BibitemOpen
  \bibfield  {author} {\bibinfo {author} {\bibfnamefont {A.}~\bibnamefont {Beck}}, \bibinfo {author} {\bibfnamefont {D.}~\bibnamefont {Flad}},\ and\ \bibinfo {author} {\bibfnamefont {C.-D.}\ \bibnamefont {Munz}},\ }\bibfield  {title} {\bibinfo {title} {Deep neural networks for data-driven {LES} closure models},\ }\bibfield  {journal} {\bibinfo  {journal} {Journal of Computational Physics}\ }\href {https://doi.org/https://doi.org/10.1016/j.jcp.2019.108910} {https://doi.org/10.1016/j.jcp.2019.108910} (\bibinfo {year} {2019})\BibitemShut {NoStop}%
\bibitem [{\citenamefont {Frezat}\ \emph {et~al.}(2021)\citenamefont {Frezat}, \citenamefont {Balarac}, \citenamefont {Sommer}, \citenamefont {Fablet},\ and\ \citenamefont {Lguensat}}]{frezat}%
  \BibitemOpen
  \bibfield  {author} {\bibinfo {author} {\bibfnamefont {H.}~\bibnamefont {Frezat}}, \bibinfo {author} {\bibfnamefont {G.}~\bibnamefont {Balarac}}, \bibinfo {author} {\bibfnamefont {J.~L.}\ \bibnamefont {Sommer}}, \bibinfo {author} {\bibfnamefont {R.}~\bibnamefont {Fablet}},\ and\ \bibinfo {author} {\bibfnamefont {R.}~\bibnamefont {Lguensat}},\ }\bibfield  {title} {\bibinfo {title} {Physical invariance in neural networks for subgrid-scale scalar flux modeling},\ }\bibfield  {journal} {\bibinfo  {journal} {Physical Review Fluids}\ }\href {https://doi.org/https://doi.org/10.1103/PhysRevFluids.6.024607} {https://doi.org/10.1103/PhysRevFluids.6.024607} (\bibinfo {year} {2021})\BibitemShut {NoStop}%
\bibitem [{\citenamefont {Novati}\ \emph {et~al.}(2021)\citenamefont {Novati}, \citenamefont {de~Laroussilhe},\ and\ \citenamefont {Koumoutsakos}}]{marl}%
  \BibitemOpen
  \bibfield  {author} {\bibinfo {author} {\bibfnamefont {G.}~\bibnamefont {Novati}}, \bibinfo {author} {\bibfnamefont {H.~L.}\ \bibnamefont {de~Laroussilhe}},\ and\ \bibinfo {author} {\bibfnamefont {P.}~\bibnamefont {Koumoutsakos}},\ }\bibfield  {title} {\bibinfo {title} {Automating turbulence modelling by multi-agent reinforcement learning},\ }\bibfield  {journal} {\bibinfo  {journal} {Nature Machine Intelligence}\ }\href {https://doi.org/10.1038/s42256-020-00272-0} {10.1038/s42256-020-00272-0} (\bibinfo {year} {2021})\BibitemShut {NoStop}%
\bibitem [{\citenamefont {Kurz}\ \emph {et~al.}(2023)\citenamefont {Kurz}, \citenamefont {Offenhauser},\ and\ \citenamefont {Beck}}]{mkurz}%
  \BibitemOpen
  \bibfield  {author} {\bibinfo {author} {\bibfnamefont {M.}~\bibnamefont {Kurz}}, \bibinfo {author} {\bibfnamefont {P.}~\bibnamefont {Offenhauser}},\ and\ \bibinfo {author} {\bibfnamefont {A.}~\bibnamefont {Beck}},\ }\bibfield  {title} {\bibinfo {title} {Deep reinforcement learning for turbulence modeling in large eddy simulations},\ }\bibfield  {journal} {\bibinfo  {journal} {International Journal of Heat and Fluid Flow}\ }\href {https://doi.org/10.1016/j.ijheatfluidflow.2022.109094} {10.1016/j.ijheatfluidflow.2022.109094} (\bibinfo {year} {2023})\BibitemShut {NoStop}%
\bibitem [{\citenamefont {Biferale}(2003)}]{biferale2003shell}%
  \BibitemOpen
  \bibfield  {author} {\bibinfo {author} {\bibfnamefont {L.}~\bibnamefont {Biferale}},\ }\bibfield  {title} {\bibinfo {title} {Shell models of energy cascade in turbulence},\ }\href {https://doi.org/10.1146/annurev.fluid.35.101101.161122} {\bibfield  {journal} {\bibinfo  {journal} {Annual Review of Fluid Mechanics}\ }\textbf {\bibinfo {volume} {35}},\ \bibinfo {pages} {441} (\bibinfo {year} {2003})}\BibitemShut {NoStop}%
\bibitem [{\citenamefont {L’vov}\ \emph {et~al.}(1998)\citenamefont {L’vov}, \citenamefont {Podivilov}, \citenamefont {Pomyalov}, \citenamefont {Procaccia},\ and\ \citenamefont {Vandembroucq}}]{LvovUnknownTitle1998}%
  \BibitemOpen
  \bibfield  {author} {\bibinfo {author} {\bibfnamefont {V.~S.}\ \bibnamefont {L’vov}}, \bibinfo {author} {\bibfnamefont {E.}~\bibnamefont {Podivilov}}, \bibinfo {author} {\bibfnamefont {A.}~\bibnamefont {Pomyalov}}, \bibinfo {author} {\bibfnamefont {I.}~\bibnamefont {Procaccia}},\ and\ \bibinfo {author} {\bibfnamefont {D.}~\bibnamefont {Vandembroucq}},\ }\bibfield  {title} {\bibinfo {title} {Improved shell model of turbulence},\ }\href {https://doi.org/10.1103/physreve.58.1811} {\bibfield  {journal} {\bibinfo  {journal} {Physical Review E}\ }\textbf {\bibinfo {volume} {58}},\ \bibinfo {pages} {1811} (\bibinfo {year} {1998})}\BibitemShut {NoStop}%
\bibitem [{\citenamefont {Hattori}\ \emph {et~al.}(2004)\citenamefont {Hattori}, \citenamefont {Rubinstein},\ and\ \citenamefont {Ishizawa}}]{ROT_SM}%
  \BibitemOpen
  \bibfield  {author} {\bibinfo {author} {\bibfnamefont {Y.}~\bibnamefont {Hattori}}, \bibinfo {author} {\bibfnamefont {R.}~\bibnamefont {Rubinstein}},\ and\ \bibinfo {author} {\bibfnamefont {A.}~\bibnamefont {Ishizawa}},\ }\bibfield  {title} {\bibinfo {title} {Shell model for rotating turbulence},\ }\href {https://doi.org/10.1103/PhysRevE.70.046311} {\bibfield  {journal} {\bibinfo  {journal} {Phys. Rev. E}\ }\textbf {\bibinfo {volume} {70}},\ \bibinfo {pages} {046311} (\bibinfo {year} {2004})}\BibitemShut {NoStop}%
\bibitem [{\citenamefont {Mingshun}\ and\ \citenamefont {Shida}(1997)}]{THERMAL_SM}%
  \BibitemOpen
  \bibfield  {author} {\bibinfo {author} {\bibfnamefont {J.}~\bibnamefont {Mingshun}}\ and\ \bibinfo {author} {\bibfnamefont {L.}~\bibnamefont {Shida}},\ }\bibfield  {title} {\bibinfo {title} {Scaling behavior of velocity and temperature in a shell model for thermal convective turbulence},\ }\href {https://doi.org/10.1103/PhysRevE.56.441} {\bibfield  {journal} {\bibinfo  {journal} {Phys. Rev. E}\ }\textbf {\bibinfo {volume} {56}},\ \bibinfo {pages} {441} (\bibinfo {year} {1997})}\BibitemShut {NoStop}%
\bibitem [{\citenamefont {Wacks}\ and\ \citenamefont {Barenghi}(2011)}]{SUPERFLUID_SM}%
  \BibitemOpen
  \bibfield  {author} {\bibinfo {author} {\bibfnamefont {D.~H.}\ \bibnamefont {Wacks}}\ and\ \bibinfo {author} {\bibfnamefont {C.~F.}\ \bibnamefont {Barenghi}},\ }\bibfield  {title} {\bibinfo {title} {Shell model of superfluid turbulence},\ }\href {https://doi.org/10.1103/PhysRevB.84.184505} {\bibfield  {journal} {\bibinfo  {journal} {Phys. Rev. B}\ }\textbf {\bibinfo {volume} {84}},\ \bibinfo {pages} {184505} (\bibinfo {year} {2011})}\BibitemShut {NoStop}%
\bibitem [{\citenamefont {Plunian}\ \emph {et~al.}(2013)\citenamefont {Plunian}, \citenamefont {Stepanov},\ and\ \citenamefont {Frick}}]{MHD_SM}%
  \BibitemOpen
  \bibfield  {author} {\bibinfo {author} {\bibfnamefont {F.}~\bibnamefont {Plunian}}, \bibinfo {author} {\bibfnamefont {R.}~\bibnamefont {Stepanov}},\ and\ \bibinfo {author} {\bibfnamefont {P.}~\bibnamefont {Frick}},\ }\bibfield  {title} {\bibinfo {title} {Shell models of magnetohydrodynamic turbulence},\ }\href {https://doi.org/https://doi.org/10.1016/j.physrep.2012.09.001} {\bibfield  {journal} {\bibinfo  {journal} {Physics Reports}\ }\textbf {\bibinfo {volume} {523}},\ \bibinfo {pages} {1} (\bibinfo {year} {2013})},\ \bibinfo {note} {shell Models of Magnetohydrodynamic Turbulence}\BibitemShut {NoStop}%
\bibitem [{\citenamefont {Benzi}\ \emph {et~al.}(1996)\citenamefont {Benzi}, \citenamefont {Biferale}, \citenamefont {Kerr},\ and\ \citenamefont {Trovatore}}]{HELIC_SM}%
  \BibitemOpen
  \bibfield  {author} {\bibinfo {author} {\bibfnamefont {R.}~\bibnamefont {Benzi}}, \bibinfo {author} {\bibfnamefont {L.}~\bibnamefont {Biferale}}, \bibinfo {author} {\bibfnamefont {R.~M.}\ \bibnamefont {Kerr}},\ and\ \bibinfo {author} {\bibfnamefont {E.}~\bibnamefont {Trovatore}},\ }\bibfield  {title} {\bibinfo {title} {Helical shell models for three-dimensional turbulence},\ }\href {https://doi.org/10.1103/PhysRevE.53.3541} {\bibfield  {journal} {\bibinfo  {journal} {Phys. Rev. E}\ }\textbf {\bibinfo {volume} {53}},\ \bibinfo {pages} {3541} (\bibinfo {year} {1996})}\BibitemShut {NoStop}%
\bibitem [{\citenamefont {Jensen}\ \emph {et~al.}(1992)\citenamefont {Jensen}, \citenamefont {Paladin},\ and\ \citenamefont {Vulpiani}}]{PASSIVE_SC_SM}%
  \BibitemOpen
  \bibfield  {author} {\bibinfo {author} {\bibfnamefont {M.~H.}\ \bibnamefont {Jensen}}, \bibinfo {author} {\bibfnamefont {G.}~\bibnamefont {Paladin}},\ and\ \bibinfo {author} {\bibfnamefont {A.}~\bibnamefont {Vulpiani}},\ }\bibfield  {title} {\bibinfo {title} {Shell model for turbulent advection of passive-scalar fields},\ }\href {https://doi.org/10.1103/PhysRevA.45.7214} {\bibfield  {journal} {\bibinfo  {journal} {Phys. Rev. A}\ }\textbf {\bibinfo {volume} {45}},\ \bibinfo {pages} {7214} (\bibinfo {year} {1992})}\BibitemShut {NoStop}%
\bibitem [{\citenamefont {Mailybaev}(2016)}]{Mailybaev_2016}%
  \BibitemOpen
  \bibfield  {author} {\bibinfo {author} {\bibfnamefont {A.~A.}\ \bibnamefont {Mailybaev}},\ }\bibfield  {title} {\bibinfo {title} {Spontaneously stochastic solutions in one-dimensional inviscid systems},\ }\href {https://doi.org/10.1088/0951-7715/29/8/2238} {\bibfield  {journal} {\bibinfo  {journal} {Nonlinearity}\ }\textbf {\bibinfo {volume} {29}},\ \bibinfo {pages} {2238} (\bibinfo {year} {2016})}\BibitemShut {NoStop}%
\bibitem [{\citenamefont {Bandak}\ \emph {et~al.}(2024)\citenamefont {Bandak}, \citenamefont {Mailybaev}, \citenamefont {Eyink},\ and\ \citenamefont {Goldenfeld}}]{bandak24}%
  \BibitemOpen
  \bibfield  {author} {\bibinfo {author} {\bibfnamefont {D.}~\bibnamefont {Bandak}}, \bibinfo {author} {\bibfnamefont {A.~A.}\ \bibnamefont {Mailybaev}}, \bibinfo {author} {\bibfnamefont {G.~L.}\ \bibnamefont {Eyink}},\ and\ \bibinfo {author} {\bibfnamefont {N.}~\bibnamefont {Goldenfeld}},\ }\bibfield  {title} {\bibinfo {title} {Spontaneous stochasticity amplifies even thermal noise to the largest scales of turbulence in a few eddy turnover times},\ }\href {https://doi.org/10.1103/PhysRevLett.132.104002} {\bibfield  {journal} {\bibinfo  {journal} {Phys. Rev. Lett.}\ }\textbf {\bibinfo {volume} {132}},\ \bibinfo {pages} {104002} (\bibinfo {year} {2024})}\BibitemShut {NoStop}%
\bibitem [{\citenamefont {Constantin}\ \emph {et~al.}(2007)\citenamefont {Constantin}, \citenamefont {Levant},\ and\ \citenamefont {Titi}}]{reg_2007}%
  \BibitemOpen
  \bibfield  {author} {\bibinfo {author} {\bibfnamefont {P.}~\bibnamefont {Constantin}}, \bibinfo {author} {\bibfnamefont {B.}~\bibnamefont {Levant}},\ and\ \bibinfo {author} {\bibfnamefont {E.~S.}\ \bibnamefont {Titi}},\ }\bibfield  {title} {\bibinfo {title} {Regularity of inviscid shell models of turbulence},\ }\href {https://doi.org/10.1103/PhysRevE.75.016304} {\bibfield  {journal} {\bibinfo  {journal} {Phys. Rev. E}\ }\textbf {\bibinfo {volume} {75}},\ \bibinfo {pages} {016304} (\bibinfo {year} {2007})}\BibitemShut {NoStop}%
\bibitem [{\citenamefont {Daumont}\ \emph {et~al.}(2000)\citenamefont {Daumont}, \citenamefont {Dombre},\ and\ \citenamefont {Gilson}}]{instanton}%
  \BibitemOpen
  \bibfield  {author} {\bibinfo {author} {\bibfnamefont {I.}~\bibnamefont {Daumont}}, \bibinfo {author} {\bibfnamefont {T.}~\bibnamefont {Dombre}},\ and\ \bibinfo {author} {\bibfnamefont {J.-L.}\ \bibnamefont {Gilson}},\ }\bibfield  {title} {\bibinfo {title} {Instanton calculus in shell models of turbulence},\ }\href {https://doi.org/10.1103/PhysRevE.62.3592} {\bibfield  {journal} {\bibinfo  {journal} {Phys. Rev. E}\ }\textbf {\bibinfo {volume} {62}},\ \bibinfo {pages} {3592} (\bibinfo {year} {2000})}\BibitemShut {NoStop}%
\bibitem [{\citenamefont {Vinuesa}\ and\ \citenamefont {Brunton}(2022)}]{vinuesa2022enhancing}%
  \BibitemOpen
  \bibfield  {author} {\bibinfo {author} {\bibfnamefont {R.}~\bibnamefont {Vinuesa}}\ and\ \bibinfo {author} {\bibfnamefont {S.~L.}\ \bibnamefont {Brunton}},\ }\bibfield  {title} {\bibinfo {title} {Enhancing computational fluid dynamics with machine learning},\ }\href {https://doi.org/10.1038/s43588-022-00264-7} {\bibfield  {journal} {\bibinfo  {journal} {Nature Computational Science}\ }\textbf {\bibinfo {volume} {2}},\ \bibinfo {pages} {358} (\bibinfo {year} {2022})}\BibitemShut {NoStop}%
\bibitem [{\citenamefont {Cho}\ \emph {et~al.}(2024)\citenamefont {Cho}, \citenamefont {Park},\ and\ \citenamefont {Choi}}]{Cho_Park_Choi_2024}%
  \BibitemOpen
  \bibfield  {author} {\bibinfo {author} {\bibfnamefont {C.}~\bibnamefont {Cho}}, \bibinfo {author} {\bibfnamefont {J.}~\bibnamefont {Park}},\ and\ \bibinfo {author} {\bibfnamefont {H.}~\bibnamefont {Choi}},\ }\bibfield  {title} {\bibinfo {title} {A recursive neural-network-based subgrid-scale model for large eddy simulation: application to homogeneous isotropic turbulence},\ }\href {https://doi.org/10.1017/jfm.2024.992} {\bibfield  {journal} {\bibinfo  {journal} {Journal of Fluid Mechanics}\ }\textbf {\bibinfo {volume} {1000}},\ \bibinfo {pages} {A76} (\bibinfo {year} {2024})}\BibitemShut {NoStop}%
\bibitem [{\citenamefont {Duraisamy}(2019)}]{prf_perspective}%
  \BibitemOpen
  \bibfield  {author} {\bibinfo {author} {\bibfnamefont {K.}~\bibnamefont {Duraisamy}},\ }\bibfield  {title} {\bibinfo {title} {Perspectives on machine learning-augmented {R}eynolds-averaged and large eddy simulation models of turbulence},\ }\bibfield  {journal} {\bibinfo  {journal} {Physical Review Fluids}\ }\textbf {\bibinfo {volume} {6}},\ \href {https://doi.org/https://doi.org/10.1103/PhysRevFluids.6.050504} {https://doi.org/10.1103/PhysRevFluids.6.050504} (\bibinfo {year} {2019})\BibitemShut {NoStop}%
\bibitem [{\citenamefont {Biferale}\ \emph {et~al.}(2017)\citenamefont {Biferale}, \citenamefont {Mailybaev},\ and\ \citenamefont {Parisi}}]{BiferaleUnknownTitle2017}%
  \BibitemOpen
  \bibfield  {author} {\bibinfo {author} {\bibfnamefont {L.}~\bibnamefont {Biferale}}, \bibinfo {author} {\bibfnamefont {A.~A.}\ \bibnamefont {Mailybaev}},\ and\ \bibinfo {author} {\bibfnamefont {G.}~\bibnamefont {Parisi}},\ }\bibfield  {title} {\bibinfo {title} {Optimal subgrid scheme for shell models of turbulence},\ }\bibfield  {journal} {\bibinfo  {journal} {Physical Review E}\ }\textbf {\bibinfo {volume} {95}},\ \href {https://doi.org/10.1103/physreve.95.043108} {10.1103/physreve.95.043108} (\bibinfo {year} {2017})\BibitemShut {NoStop}%
\bibitem [{\citenamefont {Ortali}\ \emph {et~al.}(2022)\citenamefont {Ortali}, \citenamefont {Corbetta}, \citenamefont {Rozza},\ and\ \citenamefont {Toschi}}]{OrtaliUnknownTitle2022}%
  \BibitemOpen
  \bibfield  {author} {\bibinfo {author} {\bibfnamefont {G.}~\bibnamefont {Ortali}}, \bibinfo {author} {\bibfnamefont {A.}~\bibnamefont {Corbetta}}, \bibinfo {author} {\bibfnamefont {G.}~\bibnamefont {Rozza}},\ and\ \bibinfo {author} {\bibfnamefont {F.}~\bibnamefont {Toschi}},\ }\bibfield  {title} {\bibinfo {title} {Numerical proof of shell model turbulence closure},\ }\bibfield  {journal} {\bibinfo  {journal} {Physical Review Fluids}\ }\textbf {\bibinfo {volume} {7}},\ \href {https://doi.org/10.1103/physrevfluids.7.l082401} {10.1103/physrevfluids.7.l082401} (\bibinfo {year} {2022})\BibitemShut {NoStop}%
\bibitem [{\citenamefont {Lemos}\ and\ \citenamefont {Mailybaev}(2024)}]{DominguesLemosUnknownTitle2024}%
  \BibitemOpen
  \bibfield  {author} {\bibinfo {author} {\bibfnamefont {J.~D.}\ \bibnamefont {Lemos}}\ and\ \bibinfo {author} {\bibfnamefont {A.~A.}\ \bibnamefont {Mailybaev}},\ }\bibfield  {title} {\bibinfo {title} {Data-based approach for time-correlated closures of turbulence models},\ }\bibfield  {journal} {\bibinfo  {journal} {Physical Review E}\ }\textbf {\bibinfo {volume} {109}},\ \href {https://doi.org/10.1103/physreve.109.025101} {10.1103/physreve.109.025101} (\bibinfo {year} {2024})\BibitemShut {NoStop}%
\bibitem [{\citenamefont {Um}\ \emph {et~al.}(2020)\citenamefont {Um}, \citenamefont {Brand}, \citenamefont {Fei}, \citenamefont {Holl},\ and\ \citenamefont {Thuerey}}]{um2020sol}%
  \BibitemOpen
  \bibfield  {author} {\bibinfo {author} {\bibfnamefont {K.}~\bibnamefont {Um}}, \bibinfo {author} {\bibfnamefont {R.}~\bibnamefont {Brand}}, \bibinfo {author} {\bibfnamefont {Y.}~\bibnamefont {Fei}}, \bibinfo {author} {\bibfnamefont {P.}~\bibnamefont {Holl}},\ and\ \bibinfo {author} {\bibfnamefont {N.}~\bibnamefont {Thuerey}},\ }\bibfield  {title} {\bibinfo {title} {Solver-in-the-loop: Learning from differentiable physics to interact with iterative pde-solvers},\ }\href@noop {} {\bibfield  {journal} {\bibinfo  {journal} {Advances in Neural Information Processing Systems}\ } (\bibinfo {year} {2020})}\BibitemShut {NoStop}%
\bibitem [{\citenamefont {List}\ \emph {et~al.}(2024)\citenamefont {List}, \citenamefont {Chen}, \citenamefont {Bali},\ and\ \citenamefont {Thuerey}}]{list2024temporal}%
  \BibitemOpen
  \bibfield  {author} {\bibinfo {author} {\bibfnamefont {B.}~\bibnamefont {List}}, \bibinfo {author} {\bibfnamefont {L.-W.}\ \bibnamefont {Chen}}, \bibinfo {author} {\bibfnamefont {K.}~\bibnamefont {Bali}},\ and\ \bibinfo {author} {\bibfnamefont {N.}~\bibnamefont {Thuerey}},\ }\href {http://arxiv.org/abs/2402.12971} {\bibinfo {title} {How temporal unrolling supports neural physics simulators}} (\bibinfo {year} {2024}),\ \Eprint {https://arxiv.org/abs/2402.12971} {arXiv:2402.12971 [cs.LG]} \BibitemShut {NoStop}%
\bibitem [{\citenamefont {Fukushima}(1980)}]{Fukushima_1980}%
  \BibitemOpen
  \bibfield  {author} {\bibinfo {author} {\bibfnamefont {K.}~\bibnamefont {Fukushima}},\ }\bibfield  {title} {\bibinfo {title} {Neocognitron: A self-organizing neural network model for a mechanism of pattern recognition unaffected by shift in position},\ }\href {https://doi.org/10.1007/bf00344251} {\bibfield  {journal} {\bibinfo  {journal} {Biological Cybernetics}\ }\textbf {\bibinfo {volume} {36}},\ \bibinfo {pages} {193–202} (\bibinfo {year} {1980})}\BibitemShut {NoStop}%
\bibitem [{\citenamefont {Sirignano}\ \emph {et~al.}(2020)\citenamefont {Sirignano}, \citenamefont {MacArt},\ and\ \citenamefont {Freund}}]{SirignanoUnknownTitle2020}%
  \BibitemOpen
  \bibfield  {author} {\bibinfo {author} {\bibfnamefont {J.}~\bibnamefont {Sirignano}}, \bibinfo {author} {\bibfnamefont {J.~F.}\ \bibnamefont {MacArt}},\ and\ \bibinfo {author} {\bibfnamefont {J.~B.}\ \bibnamefont {Freund}},\ }\bibfield  {title} {\bibinfo {title} {{DPM}: A deep learning {PDE} augmentation method with application to large-eddy simulation},\ }\href {https://doi.org/10.1016/j.jcp.2020.109811} {\bibfield  {journal} {\bibinfo  {journal} {Journal of Computational Physics}\ }\textbf {\bibinfo {volume} {423}},\ \bibinfo {pages} {109811} (\bibinfo {year} {2020})}\BibitemShut {NoStop}%
\bibitem [{\citenamefont {Shankar}\ \emph {et~al.}(2023)\citenamefont {Shankar}, \citenamefont {Puri}, \citenamefont {Balakrishnan}, \citenamefont {Maulik},\ and\ \citenamefont {Viswanathan}}]{shankar_burgers}%
  \BibitemOpen
  \bibfield  {author} {\bibinfo {author} {\bibfnamefont {V.}~\bibnamefont {Shankar}}, \bibinfo {author} {\bibfnamefont {V.}~\bibnamefont {Puri}}, \bibinfo {author} {\bibfnamefont {R.}~\bibnamefont {Balakrishnan}}, \bibinfo {author} {\bibfnamefont {R.}~\bibnamefont {Maulik}},\ and\ \bibinfo {author} {\bibfnamefont {V.}~\bibnamefont {Viswanathan}},\ }\bibfield  {title} {\bibinfo {title} {Differentiable physics-enabled closure modeling for {B}urgers’ turbulence},\ }\href {https://doi.org/10.1088/2632-2153/acb19c} {\bibfield  {journal} {\bibinfo  {journal} {Machine Learning: Science and Technology}\ }\textbf {\bibinfo {volume} {4}},\ \bibinfo {pages} {015017} (\bibinfo {year} {2023})}\BibitemShut {NoStop}%
\bibitem [{\citenamefont {Shankara}\ \emph {et~al.}(2024)\citenamefont {Shankara}, \citenamefont {Chakrabortya}, \citenamefont {Viswanathana},\ and\ \citenamefont {Maulik}}]{shankar_hit}%
  \BibitemOpen
  \bibfield  {author} {\bibinfo {author} {\bibfnamefont {V.}~\bibnamefont {Shankara}}, \bibinfo {author} {\bibfnamefont {D.}~\bibnamefont {Chakrabortya}}, \bibinfo {author} {\bibfnamefont {V.}~\bibnamefont {Viswanathana}},\ and\ \bibinfo {author} {\bibfnamefont {R.}~\bibnamefont {Maulik}},\ }\bibfield  {title} {\bibinfo {title} {Differentiable {T}urbulence: Closure as a {PDE}-constrained optimization},\ }\bibfield  {journal} {\bibinfo  {journal} {arXiv}\ }\href {https://doi.org/10.48550/arXiv.2307.03683} {10.48550/arXiv.2307.03683} (\bibinfo {year} {2024})\BibitemShut {NoStop}%
\bibitem [{\citenamefont {Cheng}\ and\ \citenamefont {Titterington}(1994)}]{nns}%
  \BibitemOpen
  \bibfield  {author} {\bibinfo {author} {\bibfnamefont {B.}~\bibnamefont {Cheng}}\ and\ \bibinfo {author} {\bibfnamefont {D.~M.}\ \bibnamefont {Titterington}},\ }\bibfield  {title} {\bibinfo {title} {Neural networks: A review from a statistical perspective},\ }\bibfield  {journal} {\bibinfo  {journal} {Statistical Science}\ }\href {https://doi.org/10.1214/ss/1177010638} {10.1214/ss/1177010638} (\bibinfo {year} {1994})\BibitemShut {NoStop}%
\bibitem [{\citenamefont {Kolmogorov}(1991)}]{TikhomirovUnknownTitle1991}%
  \BibitemOpen
  \bibfield  {author} {\bibinfo {author} {\bibfnamefont {A.~N.}\ \bibnamefont {Kolmogorov}},\ }\bibfield  {title} {\bibinfo {title} {The local structure of turbulence in incompressible viscous fluid for very large {R}eynolds numbers},\ }\href {https://doi.org/10.1098/rspa.1991.0075} {\bibfield  {journal} {\bibinfo  {journal} {Proc. Math. Phys. Eng. Sci}\ }\textbf {\bibinfo {volume} {434}},\ \bibinfo {pages} {9} (\bibinfo {year} {1991})}\BibitemShut {NoStop}%
\bibitem [{\citenamefont {She}\ and\ \citenamefont {Leveque}(1994)}]{SheUnknownTitle1994}%
  \BibitemOpen
  \bibfield  {author} {\bibinfo {author} {\bibfnamefont {Z.-S.}\ \bibnamefont {She}}\ and\ \bibinfo {author} {\bibfnamefont {E.}~\bibnamefont {Leveque}},\ }\bibfield  {title} {\bibinfo {title} {Universal scaling laws in fully developed turbulence},\ }\href {https://doi.org/10.1103/physrevlett.72.336} {\bibfield  {journal} {\bibinfo  {journal} {Physical Review Letters}\ }\textbf {\bibinfo {volume} {72}},\ \bibinfo {pages} {336} (\bibinfo {year} {1994})}\BibitemShut {NoStop}%
\bibitem [{\citenamefont {Hochreiter}\ and\ \citenamefont {Schmidhuber}(1997)}]{hochreiter1997lstm}%
  \BibitemOpen
  \bibfield  {author} {\bibinfo {author} {\bibfnamefont {S.}~\bibnamefont {Hochreiter}}\ and\ \bibinfo {author} {\bibfnamefont {J.}~\bibnamefont {Schmidhuber}},\ }\bibfield  {title} {\bibinfo {title} {Long short-term memory},\ }\href {https://doi.org/https://doi.org/10.1162/neco.1997.9.8.173} {\bibfield  {journal} {\bibinfo  {journal} {Neural Computation}\ }\textbf {\bibinfo {volume} {9}},\ \bibinfo {pages} {1735} (\bibinfo {year} {1997})}\BibitemShut {NoStop}%
\bibitem [{\citenamefont {Larsen}\ and\ \citenamefont {Shpeisman}(2019)}]{xla}%
  \BibitemOpen
  \bibfield  {author} {\bibinfo {author} {\bibfnamefont {R.~M.}\ \bibnamefont {Larsen}}\ and\ \bibinfo {author} {\bibfnamefont {T.}~\bibnamefont {Shpeisman}},\ }\href {https://research.google/pubs/tensorflow-graph-optimizations/} {\bibinfo {title} {Tensorflow graph optimizations}} (\bibinfo {year} {2019})\BibitemShut {NoStop}%
\end{thebibliography}%

\end{document}